\def\spitzer{{\sl Spitzer}}
\def\irac{{\sl IRAC}}

\def\galex{{\sl GALEX}}
\def\hst{{\sl HST}}

\documentclass[12pt,preprint]{aastex}
\usepackage{graphics}
\usepackage{epsfig}
\usepackage{longtable}

\begin{document}
\slugcomment{Draft version: \today}

\title{\spitzer\ Observations of Star Formation in the Extreme Outer Disk of M83 (NGC5236)}
\author{Hui Dong\altaffilmark{1}, Daniela Calzetti\altaffilmark{1}, Michael Regan\altaffilmark{2},
       David Thilker\altaffilmark{3}, Luciana Bianchi\altaffilmark{3}, Gerhardt R. Meurer\altaffilmark{3}, Fabian Walter\altaffilmark{4}}
\altaffiltext{1}{Department of Astronomy, LGRT-B 619E,
       University of Massachusetts, Amherst, MA~01003; hdong@astro.umass.edu, calzetti@astro.umass.edu}
       \altaffiltext{2}{Space Telescope Science Institute,
       Baltimore, MD~21218}
       \altaffiltext{3}{Department of Physics and Astronomy, John Hopkins University, Baltimore, MD~21218}
     \altaffiltext{4}{Max Planck Institute f\"ur Astronomie, Heidelberg, Germany}

\begin{abstract}
\spitzer\ \irac\ observations of two fields in the extended UV disk (XUV-disk)
of M83 have been recently obtained, $\sim$3~$R_{HII}$ away from the center
of the galaxy ($R_{HII}$=6.6 kpc). GALEX UV images have shown the two fields to host
in-situ recent star-formation. The
\irac\ images are used in conjunction with \galex\ data and new HI
imaging from The HI Nearby Galaxy Survey (THINGS) to constrain stellar masses
and ages of the UV clumps in the fields, and to relate the local recent star
formation to the reservoir of available gas. Multi-wavelength
photometry in the UV and mid--IR bands of 136 UV clumps (spatial
resolution $>$220~pc) identified in the two target fields,
together with model fitting of the stellar UV--MIR spectral energy
distributions (SED), suggest that the clumps cover a range of ages
between a few Myr and $>$1~Gyr with a median value around $\leq$ 100 Myr,
and have masses in the range 10$^3$--3$\times$10$^6$~M$_{\odot}$,
with a peak $\sim10^{4.7}M_{\bigodot}$.
The range of observed ages, for which only a small fraction
of the mass in stars appears to have formed in the past
$\sim$10~Myr, agrees with the dearth of H$\alpha$ emission
observed in these outer fields. At the location of our \irac\
fields, the HI map shows localized enhancement and clumping of atomic gas. A
comparison of the observed star formation with the gas reservoir
shows that the UV clumps follow the Schmidt--Kennicutt scaling law
of star formation, and that star formation is occurring in regions with gas densities
at approximately (within a factor of a few) the critical density value derived according to
the Toomre Q gravitational stability criterion. The significant 8~$\mu m$ excess in
several of the clumps (16\% of the total by number accounting for $\sim$67\% of the 8~$\mu$m
flux)) provides evidence for the existence of dust in these remote fields, in agreement with
results for other galaxies \citep[e.g.][]{pop03}. Furthermore, we observe a relatively
small excess of emission at 4.5 $\mu$m in the clumps (14\%$\pm$6\% by flux), which
suggests contribution from hot small grains ($\sim$ 1000K), as already observed
in other galaxies. From our data, the
outer regions of the M83 galaxy disk show evidence of a time-extended star
formation history over $\lesssim$1~Gyr, and of a moderately chemically-evolved
interstellar medium, in agreement with recent findings on the metallicity of the outer
HII regions of M83 \citep[]{gil07}.
\end{abstract}


\section{Introduction}\label{s:intr}
Star formation in the outer disks of galaxies has multiple
implications for our understanding of the formation and evolution
of disks, of the laws that govern star formation, and of the
interaction of massive stars with the rarefied interstellar medium
in those fields. Presence of star formation ensures that those
external fields are undergoing some chemical enrichment.
Radiative and mechanical feedback from massive stars
may be more efficient in a low--density environment, which may
thus play a key role in the enrichment of the pristine halo. The
star formation process in the outer disk takes place late
compared to the inner disk \citep{mun07}.
Furthermore, the low gas
density enables tests of the star formation threshold
\citep{mar01} and of the relation between gas density and star
formation rate density \citep[the Schmidt--Kennicutt
Law,][]{ken98} at the low end of the range. Overall, outer disks
provide insights into low--density conditions for the star
formation that may have characterized the early disk formation.

Deep H$\alpha$ imaging had already revealed outer disk star
formation beyond two optical radii ($R_{25}$)  in a few nearby
galaxies \citep{fer98,lel00}. Broadband observations had also
shown presence of significant numbers of B stars in the outer disk
of M31 \citep{cui01} and NGC 6822 \citep{deb03}. Unlike their
counterparts in the inner disks, the outer disk HII regions are
small, faint and isolated. Thus, the presence of such HII regions
does not represent a challenge to the notion that H$\alpha$
`edges' exist in most disks \citep{mar01}.

More recently, \galex\  has shown that the UV profile of disks
extends, with a smooth trend, well beyond the H$\alpha$ radius or
the R$_{25}$ of galaxies \citep[Thilker et al., 2007, in press,
Boissier et al., 2007, in press]{thi05,gil05}. XUV-disks are
apparently common in the present epoch -- though not ubiquitous --, as
they are found at least at a low level in about one-quarter of the
spiral galaxy population \citep[Thilker et al. 2007, in
press]{zar07}. The UV clumps in the outer disk are generally
associated with extended HI structures, and show evidence for
metal enrichment, with metallicities in the range
Z$_{\odot}$/5--Z$_{\odot}$/10 \citep{gil07}. Indeed, extended UV
disks are generally hosted within extended HI disks
\citep{thi05,gil05}. Thus, the rarified outer disk is not quiet
and star formation process is occurring stealthily in the extended
HI disk. The presence of smooth UV surface brightness profiles
does not counter the paucity of H$\alpha$ emission found in the
outer disk UV emission, since the UV probes a larger range of ages
than H$\alpha$ and at low--star--formation levels the number of
ionizing stars may be very small and stochastically absent
\citep{boi07}.
Thus, the outer disk
 star formation merits careful investigation, to understand its nature and the conditions
 under which it occurs.

The southern grand-design galaxy M83 (RA(2000): 13h37m00.9s, Dec(2000):
-29d51m57s) is a metal rich ($>Z_{\bigodot}$) Sc galaxy, viewed
almost face on \citep[i$\sim25^{o}$, ][]{sof99}. Its distance, ~4.5 Mpc
\citep{thi03,kar05}, makes the spatial resolution of \spitzer\ (2",
$\sim$44 pc) and GALEX (5", $\sim$109 pc), sufficient to isolate
large star formation complexes. The mean star formation rate per
unit area of M83 is roughly ~0.04 $M_{\bigodot}~yr^{-1}~kpc^{-2}$
and the total star formation rate (SFR) is about 5
$M_{\bigodot}~yr^{-1}$ \citep{ken98}. Therefore, M83 is relatively
active among the local normal star forming galaxies. The H$\alpha$
emission of M83 has a well--defined `edge' at the galactocentric distance
of around 6.6 kpc \citep{mar01}. M83 also has a very extended HI disk which
is about 3 times larger than the optical one (R$_{25}$). The possibility of a close
tidal encounter with its companion NGC 5253, around 1--2 Gyr ago \citep{rog74,van80} is
still controversial, but if confirmed, it could have triggered the starbursts in
both galaxies. Additional sources of interaction for this galaxy include dwarf
companion galaxies such as KK208 \citep{kar98}.

\galex\  images of M83 have revealed the presence of more than 100
UV-bright sources grouped in highly structured
complexes beyond the radius where the H${\alpha}$ surface
brightness decreases abruptly ~\citep{thi05}. However, only a few of these UV
bright regions (10\%-20\%) have H${\alpha}$
counterparts. This apparent discrepancy has been suggested to be due
to the wide range of the stellar populations' age within the outer
disk of M83 ~\citep{thi05} and/or to the stochastic nature of the
stellar IMF at the very low gas densities in these remote regions  \citep{boi07}.

New  mid--infrared images, from \spitzer\ \irac\ , of  two fields in the extreme outskirts of M83 are presented here,
and are combined with the \galex\ images and a 21~cm (HI) image from The HI Nearby
Galaxy Survey (THINGS, \citet{wal05,wal08}) to
investigate the properties of the UV--bright extended disk; we are exploiting the leverage offered
by the multi--wavelength observations to constrain ages and masses, and to infer the star
formation characteristics of the UV knots in the outer regions of M83. The targeted \irac\  fields show relatively high UV surface brightness and are associated with large scale filamentary HI structures.  The
THINGS HI image offers about 3~times better angular resolution than pre-existing HI images, and
enables us to explore in detail the scaling laws of star formation at a local level.

In section 2, we provide a brief description of our data. We
describe in section 3 the methods used to select the sample of
regions for the detailed analysis. The analysis of the photometry is  presented in
section 4 and our results are summarized in section 5.

\section{Observations and Data Reduction}\label{s:obser}
\begin{figure*}[!thb]
  \centerline{
      \epsfig{figure=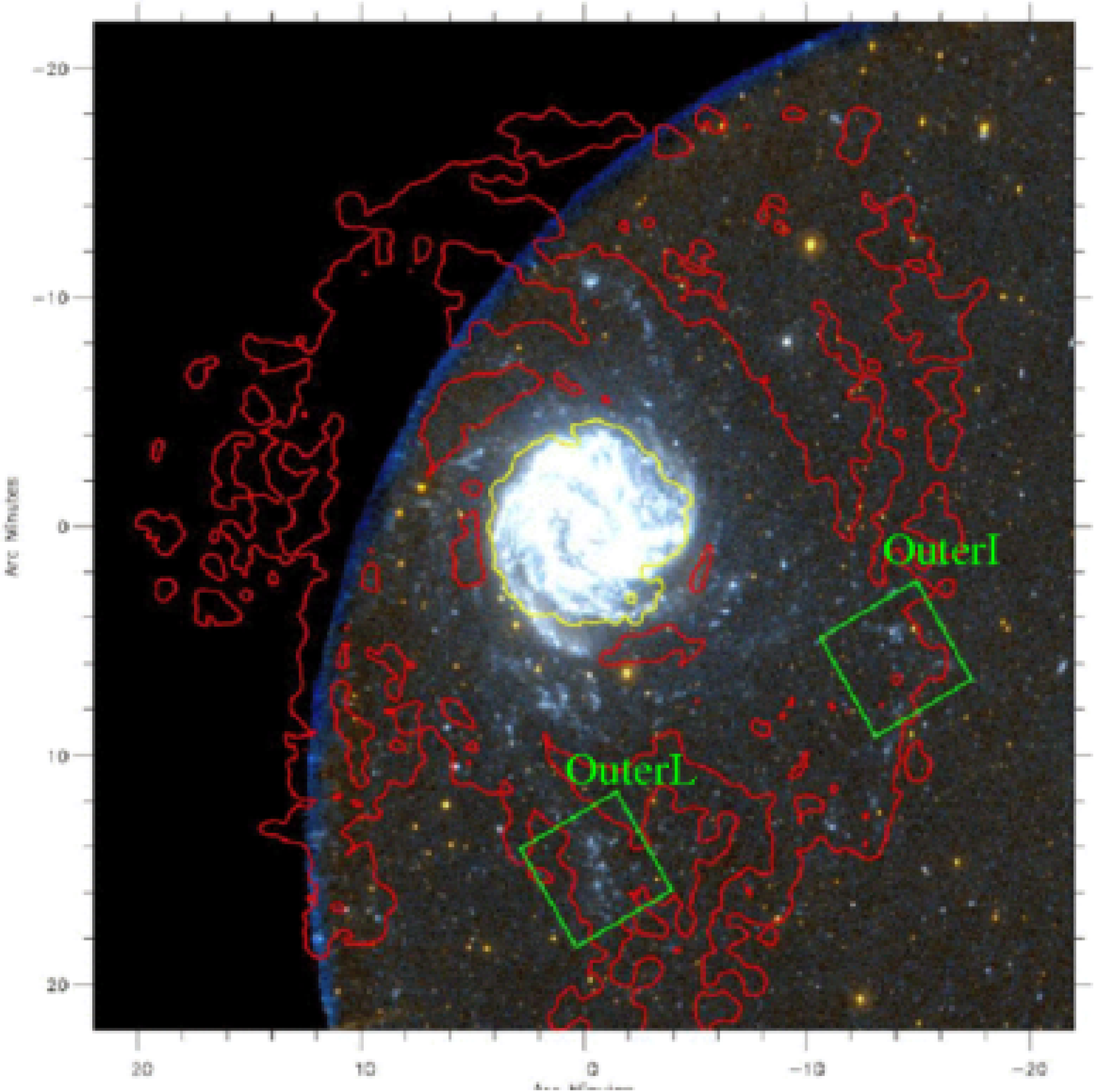,width=1\textwidth,angle=0}
    }
\caption{[adapted from \citet{thi05}]: The \galex\ image of M83 in
shown in a two--color combination, blue for the FUV channel and
orange for the NUV channel. The red contours outline the extented
HI disk \citep{til93}. The yellow contour indicates
$\Sigma_{neutral~gas}=10~M_{\bigodot}~pc^{-2}$~\citep{cro02}. The
two green squares at the right and the bottom of picture indicate
the positions and approximate size of our two \spitzer\ \irac\
fields; the name convention for the two fields is also indicated
on the figure. North is up, East is left.}
 \label{f:FUV}
 \end{figure*}

\spitzer\ images at 3.6, 4.5, 5.8 and 8.0 $\mu m$ of two UV bright
fields in the outskirts of M83 were obtained during 2005, July
using the \irac\ instrument ~\citep{faz04}. The two fields, called
OuterI and OuterL, are selected for the combination of both
UV-brightness and H$\alpha$ deficiency (see Fig.~\ref{f:FUV}) and
detailed information of these fields is listed in
Table~\ref{t:observation}. \spitzer\ \irac\ observed each field
twice with a time interval of several days to enable removal of
asteroids. The two fields were mapped with 4$\times$7 and
4$\times$6 grids, for OuterI and OuterL, respectively, with
sub-frame steps in order to reach maximum depth in fields of size
of 7.5'$\times$7.5' and 7.5'$\times$5'. The total observation time
per pixel in the center of each map is 480 second, reaching a
depth of 0.007 MJy/sr
at 3.6
$\mu$m (1 $\sigma$ detection error).
The standard \spitzer\ Infrared Nearby Galaxies Survey (SINGS)
\irac\ pipeline was employed to create the final mosaics
~\citep{ken03,reg04}.

\begin{figure*}[!thb]
  \centerline{
      \epsfig{figure=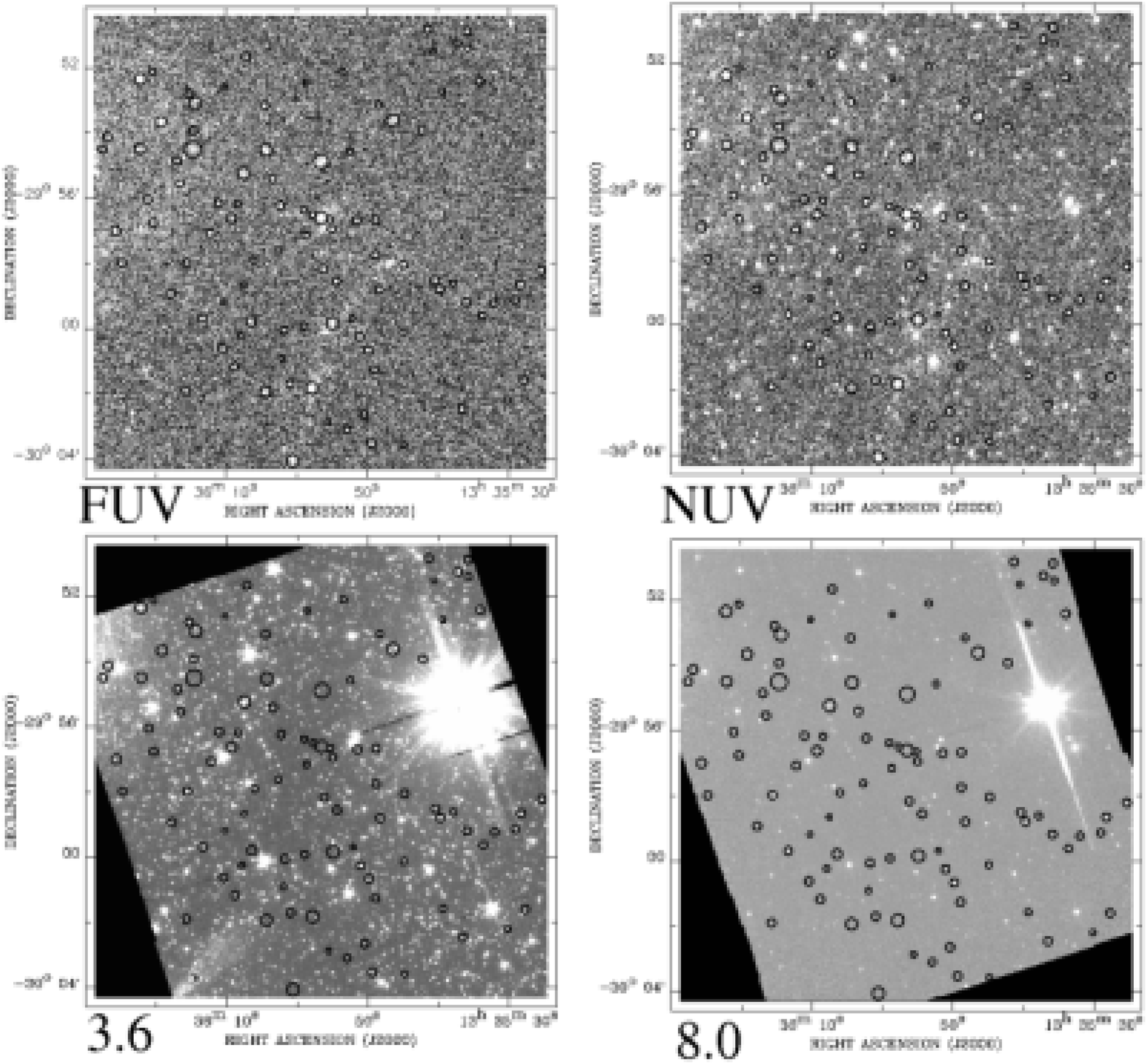,width=1.1\textwidth,angle=0}
    }
 \caption{GALEX (FUV, NUV) and \spitzer\ \irac\ (3.6 and 8.0 $\mu m$) images of the OuterI field. The black circles show the 97 sources retained in the sample
 after removal of contaminating foreground and background sources. The size of the circles
 indicates the apertures in which photometry is performed.}
 \label{f:OuterI}
 \end{figure*}

\begin{figure*}[!thb]
  \centerline{
      \epsfig{figure=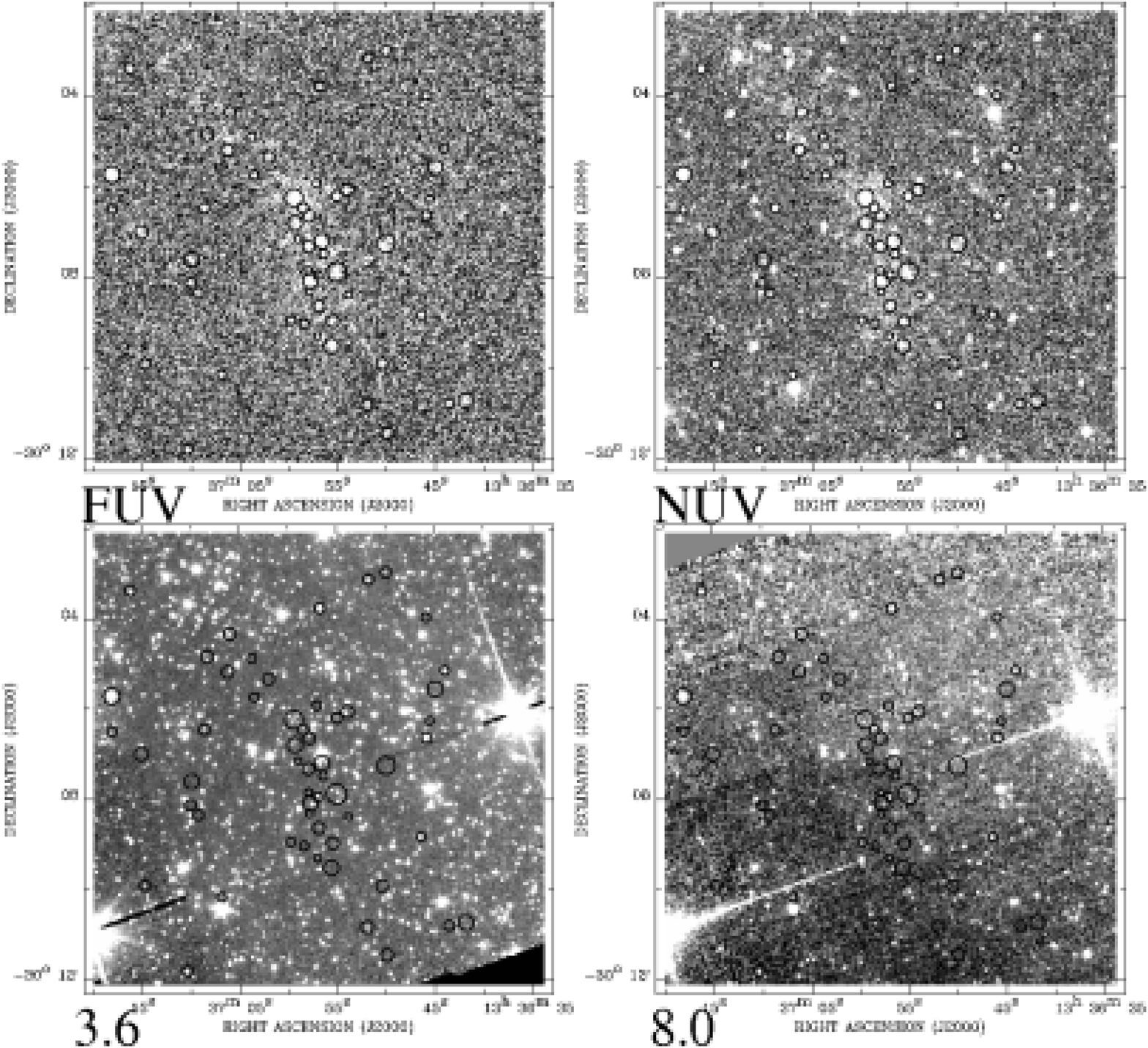,width=1.1\textwidth,angle=0}
    }
 \caption{As in Figure~\ref{f:OuterI}, now for the OuterL field. A total of 55 sources are retained
 for this field.}
 \label{f:OuterL}
 \end{figure*}

M83 was observed by \galex\ in 2003 in both the far--UV (FUV, $\lambda\sim$1529~\AA)
 and the near--UV (NUV, $\lambda\sim$2312~\AA),
as part of the Nearby Galaxy Survery \citep[NGS;][]{bia03,mar05}.
The total exposure time is 1352 s in each filter and the Point Spread Function has
a FWHM=4.6$^{\prime\prime}$. The detailed description of the \galex\
observations is provided in ~\citet{thi05}. The  \galex\ images were aligned,
registered, and cropped to match the Field-of-Views (FOVs) of the \irac\ images of
OuterI and OuterL, respectively, (Fig.~\ref{f:FUV}) and re--sampled to the same pixel size
 (0.75"/pixel for OuterI and 1"/pixel for OuterL). The \galex\ and \spitzer\  images of OuterI
 and OuterL  are depicted in Fig.~\ref{f:OuterI} and
Fig.~\ref{f:OuterL}, respectively .

The 21~cm HI image of M83, from THINGS \citep{wal05,wal08}, was reduced and calibrated
according to standard procedures developed for the THINGS project \citep[e.g.][]{wal05,ken07},
with the addition of a careful blanking in the data cube to improve sensitivity in the outer
regions of the map. Furthermore, the map has been corrected for the primary
beam attenuation, as the two \irac\ fields lay at the edges of the half-power point
of VLA primary beam ($\sim$30' diameter). The final beamsize is
15$^{\prime\prime}\times$11$^{\prime\prime}$. The final HI
image has also been aligned, registered, and cropped to match both the OuterI and OuterL FOVs.

Images from the 2MASS survey \citep{skr06} at J, H, and K$_s$ were also used in the present
analysis to provide a sanity check on our results. The 2MASS images are at least 2 orders of
magnitude less deep than our IRAC images, and the resulting uncertainty on the photometry
of our sources (which are very faint in the 2MASS fields) prevents the effective use of the 2MASS
data in our fitting routines (see below). However, the 2MASS data can still provide consistency
checks, and they will be used as such. The 2MASS images corresponding to the locations of
our fields where retrieved and aligned to both OuterI and OuterL.

\begin{deluxetable}{ccccc}
  \tabletypesize{\footnotesize}
  \tablecaption{\spitzer\ \irac\ observations}
  \tablewidth{0pt}
  \tablehead{
  \colhead{Target} &
  \colhead{Field Center Position} &
  \colhead{Galactocent.} &
  \colhead{\irac\ Map's}  &
  \colhead{exposure}  \\
  \noalign{\smallskip}
  \colhead{Field} &
  \colhead{(J2000)} &
  \colhead{Distance} &
  \colhead{core size} &
  \colhead{time(s)$\tablenotemark{a}$}
  }
  \startdata
  M83 OuterI & 13:35:56.02 -29:57:23.0 & 15.06' (19.71~kpc=2.96 $R_{HII}$) &
7.5'$\times$7.5' & 3000 \\
  M83 OuterL & 13:36:56.90 -30:06:41.7 & 14.77' (19.33~kpc=2.91 $R_{HII}$) &
7.5'$\times$5.0' & 2573 \\
  \enddata
\tablenotetext{a}{For the central map locations, the effective
exposure is 480 second}
 \label{t:observation}
 \end{deluxetable}

\section{Source Identification and Selection}\label{s:sample}
OuterI and OuterL contain a few hundred UV sources, which can be, in
addition to  star forming regions in M83, background galaxies  and
foreground stars. Identification of the nature of the sources and of the "bona fide" \irac\
counterparts to their UV emission (rather than chance
superpositions) has required extensive investigation. We developed procedures to both
exclude distant galaxies and foreground stars (often showing in
both the UV and mid-IR, M83 lies at an low ecliptic latitude of
$\sim18^{o}$), and flag UV sources that
are contaminated in the \irac\ bands by chance superposition of
physically unrelated mid-IR sources.  We applied several steps, as
discussed below, to achieve these goals.

We began constructing our sample by selecting FUV-- and
NUV--emitting regions from the \galex\ images in order to include all
potential star forming regions. The FUV is sensitive to the young
phases of SF (though not effectively instantanenous like
H$\alpha$) , while the NUV traces the slightly evolved
 stellar population (age $\sim$ 100 Myr). Sextractor ~\citep[version2.4.4,][]{ber96}
was employed to identify all 2$\sigma$ sources, yielding 226 and
133 in OuterI and OuterL, respectively (after manually combining
several adjacent and overlapping apertures in the crowded
regions). Since the instrument background and sky background of
\galex\ and \spitzer\ are much different, we did not use the
Sextractor photometry package to measure the flux of these
apertures in the \irac\ bands. Instead, the IRAF routine "phot"
was used to extract the photometry of the UV--detected sources in
all the six GALEX$+$\irac\ bands after subtracting the local
background from each of them. The radius of each region was
selected $>$5$^{\prime\prime}$ ($\sim$the \galex\ PSF,
corresponding to a physical scale $\sim$220 pc in diameter). The
same position and aperture diameter was used in all six (FUV, NUV,
3.6, 4.5, 5.8 and 8.0 $\mu$m) bands and aperture corrections are
applied to the resulting photometry. All our sources appear point--like at the
resolution of both GALEX and Spitzer/IRAC; therefore we apply to each source/bandpass
combination the point--source aperture correction appropriate for the chosen
region size. For the GALEX measurements, we derived the aperture corrections from
photometry with increasing aperture radii of point sources in the M83 field (outside the galaxy);
as a sanity check, we compared our results with the aperture corrections published by
\citet{mor07}, and found no major discrepancies with our values.
For the IRAC measurements, we referred
to the point--source aperture corrections published in the IRAC
Data Handbook\footnote{IRAC Data Handbook, v3.0, Ch. 5; http://ssc.spitzer.caltech.edu/irac/dh/ }.
Some sources are dim in several bands, and after
subtracting the local background, the flux is negative. For these
cases, a 1$\sigma$ upper limit has been set.

Uncertainties on the measurements for each source/bandpass combination
are a quadratic combination of three contributions:  variance of the local
background, photometric calibration uncertainties, and variations from potential
mis-registration of the multiwavelength images.  We find that in all cases
the variance of the local background is the predominant source of uncertainty
for our measurements, so we only consider this contribution to the uncertainties quoted in
Table~\ref{t:sou}. The local background is derived from fits of the pixel distribution of
regions of a few thousands pixels (or about 1/50th of the area of each image)
immediately surrounding each aperture, for which we derive the mode  and
the variance\footnote{The routine used for the fits, the IRAF task MSKY
written by M. Dickinson (1993, private communication), allows the user to interactively define the
interval in the pixel distribution where the mode and variance are calculated. This
ensures that robust results are obtained even in the absence of source masking.}. Each region
was also visually inspected to ensure that no systematic trends in background areas were
present. In addition, tests were run on sample apertures to verify that the mode of each local
background is not dependent on the size of the area selected for the background
measurements: we varied region sizes by up a factor of 2 (both increasing and decreasing)
without encountering for the vast majority of cases variations on excess of the measured
variances.

Contaminating foreground and background sources are eliminated through several
methods. For some nearby stars, we can identify them by
their shapes in the 3.6 $\mu$m images, i.e., by the presence of
diffraction spikes.  We also exploit  \hst\ ACS images
of the center parts of the two fields,  obtained by one of us (D.T.),  to further identify and
remove contaminated  sources from the sample; the ACS images, thanks to their
superior angular resolution, enable a quick recognition of foreground stars and background
galaxies. For the most part, we drop from our candidates list any source with strong
contamination. In those cases where the contaminating
source is physically much smaller than our measurement aperture (e.g., a distant background
galaxy) and close to its edge, we
shift and resize the measurement aperture to avoid it. In a few instances, a foreground star may lay
close to the center of our selected aperture: in these cases, we assume that the FUV
and NUV fluxes receive negligible contamination, and we only modify our \irac\
photometric measurements, by subtracting  the contaminating
contribution measured in an inset small radius aperture. In all cases, we apply the point--source
aperture correction appropriate for the aperture size selected.

Three especially bright MW stars lie within the FOVs of OuterI (to
the right) and OuterL (both right and left), see
Fig.~\ref{f:OuterI} and Fig.~\ref{f:OuterL}. They have strong
wings due to the PSF of the \spitzer\ \irac\ instrument. We reject
all apertures that fall within and close to these wings. After all
identified contaminating sources are removed from or corrected for
in our initial list, there are 166 and 114 regions remaining in
the OuterI and OuterL, respectively.

Finally, in order to remove dim foreground low mass stars more
completely, we adopt two strong selection criteria:
$log(\frac{f_{FUV}}{f_{NUV}})>-0.5$ and $f_{FUV}>5\sigma$, where
$f_{FUV}$ and $f_{NUV}$, in unit of ergs $s^{-1}$ $cm^{-2}$
$A^{-1}$,  are the flux of FUV and NUV separately and $\sigma$ is the
measurement uncertainty. Although the criteria will cause the
loss of some clusters with age larger than $\sim$ 1 Gyr (from the
prediction of Starburst99, for an instantaneous burst of star formation
with 1/5 solar abundance), it is very effective at eliminating low mass stars
which emit most of their energy in the optical and near-Infrared
band.  This `loss' of the oldest clusters will have minimal impact on this
analysis, which centers on more recent star formation events (younger than
about 1~Gyr).

The final size of our sample is 152 sources: 97 in
OuterI and 55 in OuterL. The final source list with their fluxes and
1~$\sigma$ uncertainties is listed in table~\ref{t:sou}. Although we try
our best to exclude the possible foreground stars and background
galaxies, there are still about 16 sources showing unusual high
excess 4.5, 5.8 and 8.0 $\mu$m flux which cannot be easily explained by
 stellar population synthesis models  (see section~\ref{s:dust}). These
 sources tend to be among the bright ones in our sample, and we cannot
 attribute the observed excess to photometric or background subtraction
 uncertainties. We separate these unusual sources from the other normal
sources in table~\ref{t:sou}. We consider the remaining 136
sources in our final list (although we will carry along the other
16 sources from some of the comparisons, for completeness.)

\begin{deluxetable}{rrllrrrrrrr}
  \tabletypesize{\tiny}
  \tablecaption{Source List \label{acis_source_list}}
  \tablewidth{0pt}
  \tablehead{
   \colhead{} &
   \colhead{} &
   \colhead{RA(J2000)} &
   \colhead{Dec(J2000)} &
   \colhead{}&
   \colhead{FUV$\tablenotemark{b}$} &
   \colhead{NUV$\tablenotemark{b}$} &
   \colhead{3.6$\tablenotemark{b}$} &
   \colhead{4.5$\tablenotemark{b}$} &
   \colhead{5.8$\tablenotemark{b}$} &
   \colhead{8.0$\tablenotemark{b}$} \\
   \colhead{ID} &
   \colhead{\irac\ field} &
   \colhead{degree} &
   \colhead{degree} &
   \colhead{Radius$\tablenotemark{a}$} &
   \colhead{$F_{\lambda}$ (1$\sigma$)} &
   \colhead{$F_{\lambda}$ (1$\sigma$)} &
   \colhead{$F_{\lambda}$ (1$\sigma$)} &
   \colhead{$F_{\lambda}$ (1$\sigma$)} &
   \colhead{$F_{\lambda}$ (1$\sigma$)} &
   \colhead{$F_{\lambda}$ (1$\sigma$)} \\
 }
  \startdata
  1&OuterI&13 36 22.1&-29 52 22.1&11& 4048.4( 99.2)&2700.2(38.5)& 45.6(0.65)&17.64(0.65)& 7.27(1.96)&  5.2(0.91)\\
  2&OuterI&13 36 14.3&-29 53 05.5&11& 2814.4(106.3)&1198.2(41.2)& 39.5(0.70)&19.79(0.70)& 9.85(2.09)& 19.3(0.97)\\
  3&OuterI&13 36 19.2&-29 53 40.8&10&13617.2( 92.2)&7391.8(35.8)& 15.1(0.60)& 9.48(0.60)&11.75(1.81)& 11.3(0.85)\\
  4&OuterI&13 36 26.5&-29 54 10.5& 8& 1374.7( 77.9)& 779.7(30.2)& 11.4(0.51)& 7.72(0.51)& 2.75(1.54)&  5.1(0.72)\\
  5&OuterI&13 36 17.0&-29 54 52.1& 8&  956.9( 71.0)& 418.5(27.4)& 17.3(0.47)& 8.27(0.47)& 2.67(1.40)&  3.0(0.65)\\
  6&OuterI&13 35 56.6&-29 54 54.1&14& 7945.2(127.5)&3041.1(49.3)& 31.2(0.84)&13.60(0.84)& 4.76(2.51)&  6.1(1.17)\\
  7&OuterI&13 36 07.5&-29 55 15.3&11& 3810.2(106.3)&2926.3(41.2)&156.6(0.70)&71.91(0.70)&24.98(2.09)& 29.4(0.97)\\
  8&OuterI&13 36 12.2&-29 57 06.0& 8& 2987.6( 71.0)&1245.5(27.4)& 57.2(0.47)&25.08(0.47)&11.65(1.40)&  2.8(0.65)\\
  9&OuterI&13 36 08.9&-30 01 11.3& 8& 1249.1( 71.0)& 887.7(27.4)& 20.0(0.47)& 9.11(0.47)& 6.05(1.40)&  2.7(0.65)\\
 10&OuterI&13 35 49.4&-30 03 32.5& 8& 1725.4( 71.0)& 978.8(27.4)& 80.1(0.47)&40.95(0.47)&11.70(1.40)&  8.8(0.65)\\
 11&OuterI&13 35 51.1&-30 00 16.3& 8& 1095.4( 71.0)&1015.9(27.4)&117.8(0.47)&49.38(0.47)&22.27(1.40)&  7.6(0.65)\\
 12&OuterI&13 35 48.4&-29 58 48.8& 8& 1030.5( 70.8)& 486.0(27.4)& 30.9(0.47)&15.73(0.47)& 3.88(1.39)&  5.6(0.65)\\
 13&OuterI&13 35 25.5&-29 58 13.0& 8&  878.9( 70.8)& 759.4(27.4)& 21.6(0.47)& 7.74(0.47)& 3.68(1.39)&  9.6(0.65)\\
 14&OuterI&13 35 40.5&-29 58 31.4& 8& 1195.0( 71.0)& 826.9(27.4)& 41.6(0.47)&24.10(0.47)& 8.08(1.40)& 14.3(0.65)\\
 15&OuterI&13 35 46.5&-29 53 37.8&11& 2857.7(106.3)&1765.2(41.2)& 58.4(0.70)&31.35(0.70)&11.35(2.09)& 18.2(0.97)\\
 16&OuterI&13 35 35.8&-29 51 24.8& 6&  874.6( 56.9)& 489.4(21.9)& 24.4(0.37)&11.72(0.37)& 4.58(1.12)& 15.4(0.52)\\
 17&OuterI&13 36 27.5&-29 54 30.8& 8&  588.9( 71.0)& 968.7(27.4)&171.6(0.47)&74.66(0.47)&28.05(1.40)& 11.2(0.65)\\
 18&OuterI&13 36 06.1&-29 57 55.5& 6&  759.9( 56.7)& 803.3(21.9)& 19.1(0.37)&11.11(0.37)& 3.54(1.11)& 10.0(0.52)\\
 19&OuterI&13 36 09.3&-29 56 38.2& 9&  876.8( 85.1)&1036.2(32.9)& 55.7(0.56)&31.35(0.56)&11.40(1.68)& 11.2(0.78)\\
 20&OuterI&13 35 45.0&-30 00 07.9& 6& 1026.2( 56.7)& 641.3(21.9)& 27.5(0.37)&14.54(0.37)& 4.87(1.11)&  9.6(0.52)\\
 21&OuterI&13 35 39.5&-30 01 34.8& 6&  617.0( 56.7)& 567.0(21.9)& 14.8(0.37)& 9.19(0.37)& 3.06(1.11)&  7.3(0.52)\\
 22&OuterI&13 35 50.5&-30 02 39.6& 8&  326.9( 71.0)& 742.6(27.4)& 12.3(0.47)& 5.04(0.47)& 2.75(1.40)&  1.9(0.65)\\
 23&OuterI&13 35 37.2&-29 51 15.8& 8&  439.5( 71.0)& 810.1(27.4)&243.8(0.47)&95.82(0.47)&35.00(1.40)& 11.2(0.65)\\
 24&OuterI&13 36 04.5&-29 53 10.7& 8& 1125.7( 70.8)& 637.9(27.4)&  2.8(0.47)& 0.87(0.47)& 2.83(1.39)&  0.6(0.65)\\
 25&OuterI&13 36 04.3&-29 54 32.5&11& 3680.3(106.3)&1586.4(41.2)&  0.6(0.70)& 0.44(0.70)& 2.69(2.09)&  0.7(0.97)\\
 26&OuterI&13 36 16.5&-29 55 33.7& 8& 1309.8( 71.0)& 398.3(27.4)& 12.5(0.47)& 3.68(0.47)& 2.36(1.40)&  0.4(0.65)\\
 27&OuterI&13 36 25.6&-29 57 01.2& 9& 1680.0( 85.1)& 418.5(32.9)& 82.0(0.56)&33.51(0.56)&14.49(1.68)&  1.5(0.78)\\
 28&OuterI&13 36 20.3&-29 56 46.5& 8& 1428.8( 71.0)& 381.4(27.4)&  4.2(0.47)& 3.31(0.47)& 0.68(1.40)&  0.6(0.65)\\
 29&OuterI&13 36 02.3&-29 56 15.3& 8& 1980.9( 71.0)& 789.8(27.4)&  0.0(0.47)& 0.32(0.47)& 0.85(1.40)&  0.3(0.65)\\
 30&OuterI&13 35 59.1&-29 56 23.5& 6& 2084.8( 56.7)& 594.0(21.9)&  0.2(0.37)& 0.05(0.37)& 0.71(1.11)&  1.3(0.52)\\
 31&OuterI&13 35 58.7&-29 57 10.4& 6& 2944.3( 56.9)&1340.0(21.9)&  0.8(0.37)& 1.14(0.37)& 1.53(1.12)&  0.6(0.52)\\
 32&OuterI&13 35 57.7&-29 56 31.3& 6& 1099.8( 56.9)& 489.4(21.9)&  0.3(0.37)& 0.26(0.37)& 0.71(1.12)&  0.5(0.52)\\
 33&OuterI&13 35 56.6&-29 56 37.5&11&13465.7(106.3)&6075.4(41.2)&  8.1(0.70)& 5.94(0.70)& 3.33(2.09)&  2.9(0.97)\\
 34&OuterI&13 35 55.3&-29 56 40.3& 6& 2294.8( 56.7)& 850.6(21.9)&  1.2(0.32)& 0.73(0.32)& 0.42(0.97)&  0.4(0.45)\\
 35&OuterI&13 36 06.5&-29 59 48.4& 9& 2468.0( 85.1)& 718.9(32.9)&  0.3(0.56)& 0.25(0.56)& 0.97(1.68)&  0.6(0.78)\\
 36&OuterI&13 35 56.3&-29 58 10.7& 8& 1801.2( 70.8)& 745.9(27.4)&  4.3(0.47)& 1.55(0.47)& 0.33(1.39)&  0.3(0.65)\\
 37&OuterI&13 35 54.4&-29 58 34.3& 8& 1127.9( 71.0)& 391.5(27.4)&  2.2(0.47)& 3.06(0.47)& 1.00(1.40)&  0.2(0.65)\\
 38&OuterI&13 35 55.0&-29 59 51.2&11& 4199.9(106.3)&1518.9(41.2)&  8.3(0.70)& 3.59(0.70)& 1.54(2.09)&  0.1(0.97)\\
 39&OuterI&13 36 10.5&-30 00 38.3& 8&  989.4( 71.0)& 739.2(27.4)&  4.7(0.47)& 2.41(0.47)& 1.98(1.40)&  1.0(0.65)\\
 40&OuterI&13 36 15.8&-30 01 54.9& 8& 1260.0( 71.0)& 607.5(27.4)&  6.9(0.47)& 4.21(0.47)& 1.93(1.40)&  7.0(0.65)\\
 41&OuterI&13 36 04.4&-30 01 57.0&11& 1569.5( 99.2)& 570.4(38.5)& 10.0(0.65)& 4.66(0.65)& 1.93(1.96)&  0.7(0.91)\\
 42&OuterI&13 36 01.0&-30 01 43.1& 8& 1253.5( 71.0)& 546.8(27.4)& 15.3(0.47)& 6.84(0.47)& 1.85(1.40)&  0.4(0.65)\\
 43&OuterI&13 35 57.9&-30 01 50.2&11& 5065.9(106.3)&2420.0(41.2)& 19.7(0.70)& 8.21(0.70)& 2.90(2.09)&  1.8(0.97)\\
 44&OuterI&13 36 00.7&-30 04 04.5&11& 3290.6(106.3)&1758.5(41.2)&  0.5(0.70)& 0.95(0.70)& 0.75(2.09)&  0.3(0.97)\\
 45&OuterI&13 35 36.0&-29 59 11.8& 8&  840.0( 70.8)& 421.9(27.4)&  9.1(0.47)& 3.35(0.47)& 0.57(1.39)&  2.1(0.65)\\
 46&OuterI&13 35 44.9&-29 58 03.4& 8& 1608.5( 71.0)& 567.0(27.4)&  3.6(0.47)& 1.11(0.47)& 0.33(1.40)&  0.1(0.65)\\
 47&OuterI&13 36 14.6&-29 53 56.9& 8&  346.4( 71.0)& 749.3(27.4)& 16.0(0.47)& 8.43(0.47)& 2.09(1.40)&  8.6(0.65)\\
 48&OuterI&13 36 15.3&-29 52 48.7& 8&  378.9( 71.0)& 843.8(27.4)& 25.6(0.47)&10.93(0.47)& 0.66(1.40)&  1.7(0.65)\\
 49&OuterI&13 36 010.&-29 52 37.4& 5&  287.9( 42.4)& 418.5(16.5)&  6.4(0.28)& 2.82(0.28)& 0.27(0.84)&  0.5(0.39)\\
 50&OuterI&13 35 58.6&-29 52 27.5& 5&  242.5( 42.4)& 325.4(16.5)&  3.1(0.28)& 0.91(0.28)& 0.36(0.84)&  0.1(0.39)\\
 51&OuterI&13 35 40.6&-29 51 30.9& 5&  456.8( 65.2)&1026.1(24.1)& 23.3(0.41)& 8.37(0.42)& 0.91(1.22)&  0.0(0.59)\\
 52&OuterI&13 36 07.6&-29 58 40.5& 5&  257.6( 42.4)& 371.3(16.5)& 32.1(0.28)&12.66(0.28)& 5.39(0.84)&  0.4(0.39)\\
 53&OuterI&13 36 07.9&-30 00 15.0& 5&  521.7( 49.6)& 658.2(19.2)&  4.4(0.33)& 2.33(0.33)& 1.96(0.98)&  0.4(0.45)\\
 54&OuterI&13 35 52.1&-29 59 42.2& 5&  519.6( 42.4)& 408.4(16.5)&  1.3(0.28)& 0.28(0.28)& 1.06(0.84)&  0.2(0.39)\\
 55&OuterI&13 36 07.1&-29 51 41.1& 8&  592.6( 71.0)& 358.8(27.4)&  0.9(0.47)& 0.89(0.47)& 0.85(1.40)&  0.6(0.65)\\
 56&OuterI&13 36 14.6&-29 54 31.5&15& 3788.6(141.7)&1066.6(54.8)& 16.2(0.93)&12.61(0.93)& 3.40(2.79)&  2.3(1.30)\\
 57&OuterI&13 36 24.8&-29 58 00.1& 8&  708.8( 71.0)& 315.4(27.4)&  3.8(0.47)& 0.58(0.47)& 3.75(1.40)&  3.4(0.65)\\
 58&OuterI&13 35 27.9&-30 01 36.0& 8& 1027.9( 71.0)& 331.6(27.4)& 70.2(0.47)&27.10(0.47)&11.14(1.40)&  2.1(0.65)\\
 59&OuterI&13 35 42.3&-29 53 56.5& 8&  790.9( 71.0)& 514.1(27.4)& 11.1(0.47)& 6.94(0.47)& 1.66(1.40)&  5.1(0.65)\\
 60&OuterI&13 36 20.4&-29 52 05.9& 6&  338.5( 56.7)& 281.9(21.9)&  4.3(0.37)& 3.69(0.37)& 4.58(1.12)&  0.8(0.52)\\
 61&OuterI&13 36 22.1&-29 54 30.7&10& 1530.4( 92.1)& 805.6(35.7)&  8.2(0.61)& 2.85(0.60)& 1.02(1.81)&  0.7(0.84)\\
 62&OuterI&13 36 21.1&-29 56 03.8& 8& 1021.1( 71.0)& 291.9(27.4)&  3.0(0.47)& 0.37(0.47)& 1.70(1.40)&  0.8(0.65)\\
 63&OuterI&13 36 11.1&-29 56 10.5& 8&  893.4( 71.0)& 521.4(27.4)& 13.3(0.47)& 6.85(0.47)& 1.70(1.40)&  3.8(0.65)\\
 64&OuterI&13 36 01.8&-30 00 04.5& 8&  647.3( 71.0)& 333.1(27.4)&  6.2(0.47)& 1.86(0.47)& 0.85(1.40)&  2.1(0.65)\\
 65&OuterI&13 35 36.6&-30 02 28.1& 8&  806.8( 71.0)& 211.1(27.4)& 13.7(0.47)& 6.35(0.47)& 1.35(1.40)&  2.8(0.65)\\
 66&OuterI&13 35 48.9&-29 57 45.5& 8&  806.8( 71.0)& 325.0(27.4)&  3.0(0.47)& 1.65(0.47)& 3.03(1.40)&  0.2(0.65)\\
 67&OuterI&13 35 51.5&-29 56 42.5& 8&  713.4( 57.1)& 672.0(23.3)&  8.1(0.39)& 3.60(0.39)& 5.23(1.19)&  0.5(0.53)\\
 68&OuterI&13 35 35.9&-29 50 53.2& 8&  467.2( 57.1)& 350.0(23.3)&  5.1(0.39)& 1.16(0.39)& 1.16(1.19)&  0.9(0.53)\\
 69&OuterI&13 35 41.5&-29 50 50.4& 8&  692.9( 71.0)& 251.5(27.4)&  3.9(0.47)& 2.21(0.47)& 0.85(1.40)&  0.6(0.65)\\
 70&OuterI&13 35 49.0&-30 01 16.3& 8&  542.4( 71.0)&  22.1(27.4)&  1.8(0.47)& 0.58(0.47)& 0.85(1.40)&  1.5(0.65)\\
 71&OuterI&13 35 59.0&-29 59 56.2& 7&  673.5( 63.9)& 518.4(24.7)&  4.3(0.42)& 2.29(0.42)& 5.43(1.26)&  1.7(0.58)\\
 72&OuterI&13 36 02.8&-29 57 38.2& 6&  426.0( 56.8)& 455.4(21.9)&  7.7(0.37)& 3.64(0.37)& 2.62(1.12)&  2.0(0.52)\\
 73&OuterI&13 36 02.1&-30 00 55.5& 5&  273.1( 49.6)& 358.7(19.2)& 12.2(0.33)& 5.90(0.33)& 2.74(0.98)&  2.5(0.45)\\
 74&OuterI&13 35 55.7&-30 02 52.7& 5&  296.2( 49.6)& 311.3(19.2)& 10.6(0.33)& 4.47(0.33)& 5.12(0.98)&  1.3(0.45)\\
 75&OuterI&13 35 44.9&-30 03 36.9& 5&  261.5( 49.6)& 557.0(19.2)& 15.4(0.33)&10.34(0.33)& 5.35(0.98)&  0.4(0.45)\\
 76&OuterI&13 35 53.0&-30 03 06.7& 7&  815.6( 63.9)& 436.9(24.7)&  7.8(0.42)& 6.25(0.42)& 3.09(1.26)&  2.0(0.58)\\
 77&OuterI&13 35 29.2&-29 59 08.3& 7&  643.7( 63.9)& 403.6(24.7)& 60.7(0.42)&29.28(0.42)& 8.23(1.26)&  5.0(0.58)\\
 78&OuterI&13 35 32.1&-29 59 14.1& 7&  410.1( 63.7)& 599.8(24.7)&  9.5(0.42)& 2.82(0.42)& 0.78(1.26)&  0.5(0.58)\\
 79&OuterI&13 35 33.8&-29 59 37.7& 7&  611.7( 63.7)& 388.8(24.7)& 11.8(0.42)& 7.62(0.42)& 0.78(1.26)&  2.1(0.58)\\
 80&OuterI&13 35 38.0&-29 58 37.5& 6&  605.6( 56.7)& 316.2(21.9)& 11.3(0.37)& 6.68(0.37)& 4.42(1.12)&  0.6(0.52)\\
 81&OuterI&13 35 52.5&-29 54 35.0& 6&  393.7( 56.7)& 360.2(21.9)& 11.7(0.37)& 6.44(0.37)& 4.54(1.12)&  6.4(0.52)\\
 82&OuterI&13 35 48.4&-29 53 10.1& 6&  578.0( 56.7)& 466.6(21.9)& 15.9(0.37)& 7.82(0.37)& 3.86(1.12)&  2.5(0.52)\\
 83&OuterI&13 35 53.5&-29 52 07.0& 6&  538.8( 56.7)& 392.0(21.9)& 18.0(0.37)& 9.09(0.37)& 3.19(1.12)&  5.1(0.52)\\
 84&OuterI&13 35 39.5&-29 52 43.6& 5&  541.6( 49.6)& 250.7(19.2)&  6.2(0.33)& 0.09(0.33)&14.01(0.98)&  1.7(0.45)\\
 85&OuterL&13 36 55.1&-30 05 48.7& 7& 5000.9(118.6)&2990.5(33.4)&  4.0(0.68)& 1.71(0.66)& 2.39(1.82)&  5.0(1.04)\\
 86&OuterL&13 36 59.5&-30 05 49.7&12& 8464.7(203.5)&4927.8(57.0)& 41.1(1.16)&18.92(1.13)& 8.57(3.12)&  4.5(1.78)\\
 87&OuterL&13 37 15.2&-30 06 30.4& 9& 2446.3(152.6)&1036.2(42.9)&  6.6(0.87)& 2.79(0.85)& 3.54(2.34)&  1.6(1.33)\\
 88&OuterL&13 37 07.3&-30 09 42.1& 6& 1275.1(101.8)& 594.0(28.6)&  9.8(0.58)& 4.74(0.57)& 5.17(1.56)&  1.6(0.89)\\
 89&OuterL&13 36 54.0&-30 05 40.5& 8& 1677.8(135.5)&1498.6(38.1)& 40.5(0.77)&17.87(0.75)& 7.44(2.08)&  8.1(1.18)\\
 90&OuterL&13 36 58.6&-30 06 03.4& 7& 2251.5(118.6)&1461.5(33.4)&  5.1(0.68)& 2.78(0.66)& 8.05(1.82)& 10.4(1.04)\\
 91&OuterL&13 36 46.7&-30 08 28.8& 6& 2026.3(101.8)&1407.5(28.6)& 56.0(0.58)&24.14(0.57)&12.99(1.56)&  2.4(0.89)\\
 92&OuterL&13 37 05.9&-30 03 54.2& 8&  989.4(135.7)& 907.9(38.1)& 40.5(0.77)&17.18(0.75)& 9.05(2.08)&  3.4(1.18)\\
 93&OuterL&13 36 46.0&-30 06 17.7& 7&  779.4(118.6)&1302.8(33.4)& 70.2(0.68)&32.64(0.66)&15.50(1.82)&  4.3(1.04)\\
 94&OuterL&13 36 57.3&-30 05 33.5& 6& 1489.4(101.8)& 735.8(28.6)&  2.9(0.58)& 0.44(0.57)& 0.97(1.56)&  1.1(0.89)\\
 95&OuterL&13 36 57.9&-30 06 14.6& 8& 2359.7(135.5)&1073.3(38.1)&  3.6(0.77)& 0.45(0.75)& 2.39(2.08)&  0.9(1.18)\\
 96&OuterL&13 36 59.4&-30 06 23.2&10& 2749.4(169.7)&1410.8(47.6)&  6.0(0.97)& 6.55(0.94)& 2.54(2.60)&  0.3(1.48)\\
 97&OuterL&13 36 56.7&-30 06 48.2&10& 7707.0(169.7)&4286.5(47.6)& 27.4(0.58)&11.29(0.57)& 3.00(1.56)&  1.5(0.89)\\
 98&OuterL&13 36 58.1&-30 06 55.3& 8& 2662.8(135.7)&1363.6(38.1)&  6.4(0.77)& 3.62(0.75)& 1.36(2.08)&  0.3(1.18)\\
 99&OuterL&13 36 58.0&-30 07 40.7&10& 3269.0( 48.5)&1046.3(13.6)& 17.4(0.26)& 8.01(0.27)& 2.20(0.74)&  1.4(0.43)\\
100&OuterL&13 36 54.2&-30 08 00.6& 5& 1530.6( 84.9)& 931.6(23.9)&  2.8(0.48)& 0.52(0.47)& 1.41(1.30)&  1.1(0.74)\\
101&OuterL&13 36 57.2&-30 08 14.1& 8& 2141.1(135.5)&1400.7(38.1)&  4.7(0.77)& 1.95(0.75)& 0.53(2.08)&  2.7(1.18)\\
102&OuterL&13 36 50.1&-30 06 53.2&12& 1946.2(172.1)& 860.7(40.2)&  0.9(1.00)& 0.44(0.95)& 8.05(2.65)&  0.4(1.50)\\
103&OuterL&13 36 55.8&-30 08 35.0& 8& 2048.0(135.7)& 776.3(38.1)&  1.8(0.77)& 2.06(0.75)& 1.93(2.08)&  2.0(1.18)\\
104&OuterL&13 36 50.8&-30 09 32.5& 8& 1182.0(135.5)& 300.4(38.1)&  7.2(0.77)& 2.72(0.75)& 1.50(2.08)&  2.6(1.18)\\
105&OuterL&13 36 56.0&-30 09 07.3&10& 2468.0(169.7)&1346.7(47.6)&  1.3(0.97)& 0.30(0.94)& 2.03(2.60)&  0.3(1.48)\\
106&OuterL&13 36 49.8&-30 02 37.3& 7& 1753.6(118.6)&1515.5(33.4)&  5.5(0.68)& 2.67(0.66)& 3.48(1.82)&  0.4(1.04)\\
107&OuterL&13 37 06.2&-30 04 44.2& 8& 2154.1(135.7)&2146.6(38.1)& 31.4(0.77)&12.37(0.75)& 4.23(2.08)&  0.5(1.18)\\
108&OuterL&13 37 08.7&-30 06 00.2& 7& 1872.6(118.6)&1262.3(33.4)& 10.6(0.68)& 4.56(0.66)& 2.07(1.82)&  2.8(1.04)\\
109&OuterL&13 37 03.5&-30 05 18.5& 6&  956.9(101.8)& 415.2(28.6)&  1.5(0.58)& 0.33(0.57)& 0.30(1.56)&  2.2(0.89)\\
110&OuterL&13 37 03.7&-30 04 27.1& 6&  917.9(101.8)& 266.3(28.6)&  5.4(0.58)& 2.19(0.57)& 1.06(1.56)&  0.4(0.89)\\
111&OuterL&13 36 52.4&-30 10 26.9& 8&  850.8(135.5)&1107.1(38.1)& 23.2(0.77)& 8.88(0.75)& 0.08(2.08)&  1.8(1.18)\\
112&OuterL&13 36 50.4&-30 11 04.5& 8& 1002.3(135.7)& 384.8(38.1)&  0.9(0.77)& 2.86(0.75)& 1.38(2.08)&  0.5(1.18)\\
113&OuterL&13 36 55.3&-30 07 29.4&13& 2424.7(220.8)&1691.0(62.1)& 18.0(1.18)&10.73(1.15)& 2.72(3.18)&  2.7(1.81)\\
114&OuterL&13 36 45.6&-30 05 55.4& 5&  961.2( 84.9)& 401.7(23.9)&  0.2(0.48)& 0.57(0.47)& 0.46(1.30)&  0.9(0.74)\\
115&OuterL&13 37 10.3&-30 07 39.4& 7&  956.9(118.6)& 344.3(33.4)&  5.9(0.68)& 1.80(0.66)& 0.33(1.82)&  1.5(1.04)\\
116&OuterL&13 36 45.0&-30 05 13.5&10& 1961.4(169.5)& 982.2(47.6)&  2.6(0.97)& 1.81(0.94)& 1.42(2.60)&  0.2(1.48)\\
117&OuterL&13 36 59.2&-30 06 45.4& 5& 1405.0( 84.9)&1083.5(23.9)&  0.5(0.19)& 0.30(0.19)& 0.52(0.52)&  0.0(0.30)\\
118&OuterL&13 36 51.6&-30 02 46.1& 7&  459.0(118.6)& 432.0(33.4)& 11.2(0.68)& 6.41(0.66)& 1.67(1.82)&  4.7(1.04)\\
119&OuterL&13 37 10.0&-30 07 37.7&10& 1355.2(169.6)& 517.7(47.6)&  0.3(0.97)& 0.32(0.94)& 0.99(2.60)&  0.4(1.48)\\
120&OuterL&13 36 41.9&-30 10 46.1&10& 2231.1(169.6)& 699.8(47.6)&  1.8(0.97)& 3.25(0.94)& 0.99(2.60)&  0.4(1.48)\\
121&OuterL&13 36 43.7&-30 10 50.0& 7&  919.4(118.7)& 568.6(33.4)&  3.3(0.68)& 0.84(0.66)& 0.35(1.82)&  1.3(1.04)\\
122&OuterL&13 36 56.8&-30 03 49.1& 7&  651.8(118.7)& 259.2(33.4)& 60.9(0.68)&32.67(0.66)&16.40(1.82)&  5.2(1.04)\\
123&OuterL&13 36 46.0&-30 04 01.7& 6&  651.6(101.8)& 380.8(28.6)& 20.2(0.58)& 0.21(0.56)& 2.21(1.56)&  0.3(0.89)\\
124&OuterL&13 37 02.1&-30 05 22.4& 8& 1183.4(135.6)& 322.8(38.1)&  1.8(0.77)& 0.13(0.75)& 0.62(2.08)&  0.2(1.18)\\
125&OuterL&13 37 08.4&-30 04 52.4& 8&  897.2(135.6)& 434.6(38.1)& 19.5(0.77)&10.11(0.75)& 4.07(2.08)&  2.7(1.18)\\
126&OuterL&13 37 09.3&-30 08 22.5& 7&  917.1(118.6)& 305.0(33.4)&  2.6(0.68)& 0.53(0.66)& 0.64(1.82)&  0.3(1.04)\\
127&OuterL&13 36 59.8&-30 08 58.8& 6&  660.9(101.8)& 283.3(28.6)&  2.3(0.58)& 1.43(0.57)& 3.04(1.56)&  0.4(0.89)\\
128&OuterL&13 36 58.5&-30 09 03.7& 6&  594.1(101.8)& 216.5(28.6)&  0.2(0.58)& 0.21(0.56)& 0.67(1.56)&  0.3(0.89)\\
129&OuterL&13 37 18.1&-30 06 30.9& 6&  723.0(101.8)& 392.0(28.6)& 10.3(0.58)& 4.33(0.56)& 2.84(1.56)&  0.3(0.89)\\
130&OuterL&13 36 44.1&-30 05 11.8& 6&  582.6(101.8)& 403.2(28.6)& 22.6(0.58)&14.42(0.57)& 7.12(1.56)& 10.1(0.89)\\
131&OuterL&13 37 14.8&-30 09 55.7& 7&  773.0(118.7)& 546.5(33.4)&  5.1(0.68)& 0.67(0.66)& 0.92(1.82)&  0.3(1.04)\\
132&OuterL&13 36 56.5&-30 07 31.0& 6& 1022.4(101.8)& 630.9(28.6)&  0.4(0.58)& 0.21(0.57)& 0.67(1.56)&  0.3(0.89)\\
133&OuterL&13 36 58.0&-30 07 52.9& 5&  823.0( 84.8)& 509.5(23.9)&  1.2(0.48)& 0.56(0.47)& 0.41(1.30)&  0.5(0.74)\\
134&OuterL&13 36 57.9&-30 08 20.8& 5&  635.2( 84.8)& 486.8(23.9)&  2.8(0.48)& 1.22(0.47)& 1.44(1.30)&  0.4(0.74)\\
135&OuterL&13 36 57.1&-30 07 59.0& 6&  536.5(101.8)& 873.6(28.6)& 65.8(0.58)&27.69(0.57)& 9.00(1.56)&  2.5(0.89)\\
136&OuterL&13 36 57.1&-30 09 19.8& 5&  665.3( 84.8)& 352.5(23.9)&
3.5(0.48)& 1.81(0.47)& 0.89(1.30)&  0.8(0.74) \nl \cline{1-11}
\multicolumn{11}{c}{} \\
\multicolumn{11}{c}{Sources with unusual high excess 4.5, 5.8 and
8.0 $\mu$m flux} \nl \multicolumn{10}{c}{}\\ \cline{1-11}
137&OuterI&13 36 03.4&-29 55 25.7& 8& 1420.2( 70.8)& 742.6(27.4)& 57.5(0.47)&29.98(0.47)&15.08(1.39)& 39.5(0.65)\\
138&OuterI&13 35 55.1&-29 56 57.5& 7& 3702.0( 63.9)&1937.4(24.7)& 13.7(0.42)& 7.54(0.42)& 8.18(1.26)&  5.7(0.58)\\
139&OuterI&13 36 13.4&-29 59 42.1& 8& 1474.3( 71.0)&2028.5(27.4)& 61.2(0.47)&61.33(0.47)&51.61(1.40)& 35.7(0.65)\\
140&OuterI&13 35 49.9&-30 00 40.7& 8& 5022.6( 71.0)&5299.1(27.4)& 56.0(0.47)&55.45(0.47)&50.67(1.40)& 40.0(0.65)\\
141&OuterI&13 35 28.4&-29 58 39.7& 8& 1907.3( 71.0)&1437.9(27.4)& 52.0(0.47)&21.95(0.47)&13.67(1.40)& 24.9(0.65)\\
142&OuterI&13 35 39.8&-29 58 48.3& 8& 1405.0( 71.0)&1066.6(27.4)& 75.6(0.47)&43.50(0.47)&19.21(1.40)& 65.1(0.65)\\
143&OuterI&13 36 15.6&-29 58 00.0& 8&  493.6( 71.0)& 864.1(27.4)& 93.3(0.47)&60.74(0.47)&41.36(1.40)& 68.2(0.65)\\
144&OuterI&13 35 48.9&-29 56 40.9& 8&  532.6( 71.0)& 617.7(27.4)& 24.3(0.47)&20.77(0.47)&15.79(1.40)&  9.4(0.65)\\
145&OuterI&13 35 34.1&-29 52 25.5& 8&  703.6( 71.0)& 621.0(27.4)& 61.5(0.47)&34.68(0.47)&18.38(1.40)& 20.5(0.65)\\
146&OuterI&13 36 10.3&-29 59 12.0& 5&  322.6( 42.4)& 388.2(16.5)& 31.5(0.28)&15.52(0.28)& 9.17(0.84)& 21.6(0.39)\\
147&OuterI&13 36 17.8&-29 58 57.1& 8& 1030.2( 71.0)& 223.6(27.4)& 26.8(0.47)&14.98(0.47)& 8.94(1.40)&  8.8(0.65)\\
148&OuterI&13 36 08.4&-29 56 11.9& 6&  561.8( 56.7)& 380.8(21.9)& 21.5(0.37)&12.85(0.37)& 8.66(1.12)&  6.1(0.52)\\
149&OuterI&13 35 30.3&-30 02 12.1& 5&  425.8( 49.6)& 342.5(19.2)& 15.2(0.33)& 8.91(0.33)& 5.46(0.98)& 10.2(0.45)\\
150&OuterL&13 37 16.0&-30 02 53.2& 7& 1340.1(118.9)&2028.5(33.4)& 31.0(0.68)&26.34(0.66)&21.16(1.82)& 16.5(1.04)\\
151&OuterL&13 37 18.1&-30 05 13.0&11& 8140.0(186.4)&6986.7(52.3)&176.2(1.06)&79.08(1.03)&74.38(2.87)&113.5(1.63)\\
152&OuterL&13 37 10.9&-30 11 20.1& 7&  610.5(118.6)& 853.9(33.4)& 25.3(0.68)&18.88(0.66)&19.36(1.82)&  5.7(1.04)\\
\enddata
\tablecomments{(a) Radius of the photometric aperture, in units of
arc second. (b) Fluxes are in units of
$10^{-19}~ergs~s^{-1}~cm^{-2}~A^{-1}$; the 1~$\sigma$ error bars
are indicated in parenthesis, following the flux value. For FUV
and NUV band, the flux has been corrected for the foreground Milky
Way extinction, E(B-V)=0.066 [$A_{FUV}=8.376E(B-V)$ and
$A_{NUV}=8.741E(B-V)$~\citep{wyd05}] after convolving the GALEX
spectral response curve with the Galaxy's extinction curve, using
the relation of \citet{car89} with R$_V$=3.1); the flux in the
\spitzer \irac\ bands is assumed to be extinction free.}
\label{t:sou}
\end{deluxetable}

\section{Analysis}\label{s:disc}
\subsection{Stellar populations' modelling} \label{s:theo}
The age and mass of each source is constrained by fitting the UV$+$MIR bands
with models of the spectral energy distribution (SEDs) of single--age
stellar populations. This implicitly assumes that the emission in the \irac\ bands is due to
photospheric emission from stars. This is a reasonable assumption for the 3.6 and  4.5~$\mu$m
bands \citep{pah04,cal05}. The 5.8 and 8~$\mu$m band potentially contain a significant (or dominant)
contribution from dust emission, due to Policyclic Aromatic Hydrocarbons single--photon heated
by the stellar populations; to evaluate the impact of dust emission on our stellar population fits,
we create a dust--only 8~$\mu$m image by rescaling and subtracting the 3.6~$\mu$m image from
the 8~$\mu$m one. The resulting dust--only image shows a number of sources with detected dust
emission, in most cases a weak detection, in agreement with the expectation that low--metallicity
regions  have underluminous 8~$\mu$m dust emission \citep{eng05,cal07}. However, to
prevent out fitting results from being even mildly contaminated by dust emission contributions,
we use only the two \irac\ bands at the shortest wavelengths in our fits, together with the
two GALEX bands (next section).

The theoretical SEDs are from the Starburst99 stellar population synthesis models
\citep{lei99,vaz05}.
We select instantaneous burst populations, since  we are likely to deal with single or
small groups of stellar clusters (i.e., same or similar age populations) due to the small physical
sizes of the regions sampled (100pc to 300pc). A metallicity value
of Z=0.004 is adopted for the models, to match as closely as possible the oxygen
abundances of roughly 1/10-1/5 $Z_{\bigodot}$ measured in a few of the HII regions
in these areas \citep[][]{gil07}. The Kroupa Initial Mass Function
($\Phi$(m)$\propto$$m^{-\alpha}$, with $\alpha$=1.3 for
0.1$M_{\bigodot}<M<0.5M_{\bigodot}$,  and $\alpha$=2.3 for
0.5$M_{\bigodot}<M<100M_{\bigodot}$, )\citep{kro01},
the default IMF of Starburst99, is employed. Since we include in the fits the mid-Infrared (MIR)
fluxes of each stellar complex, we use the new Padova stellar
models with full asymptotic giant branch evolution. In order to
balance the accuracy and the speed of the simulations for
different age ranges, we select different time steps for the
models: 0.05 Myr for 0.1-1 Myr, 0.1 Myr for 1-10 Myr, 0.5 Myr for
10-50 Myr, 1 Myr for 50-200 Myr, 2 Myr for 200-500 Myr, 10 Myr for
500-1000 Myr and 50 Myr for 1-2 Gyr. We convolve the
spectral response curve of the \galex\ and
\spitzer\ filter bandpasses with the simulated stellar population spectra  to
obtain the synthetic photometry in the six bands. Finally, the luminosities and the colors
from observations are compared with
the theoretical model to constrain ages and masses.

\subsection{Mass and Age distributions}\label{s:mass}
For the SED fitting procedure, we adopt the $\chi^{2}$--minimization
technique to compare the observation and the theoretical SEDs of the sources:
\begin{equation}
\chi^{2}(t,E(B-V),m,Z)=\sum_{N}\frac{(L_{obs}-A*L_{model})^2}{\sigma^2_{obs}}
\end{equation}
where N is the number of the filters available for each source
(due to the potential dust contamination of 5.8 and 8.0 $\mu$m
bands, the other four bands, FUV, NUV, 3.6 and 4.5 $\mu$m are
used, then N=4). $L_{obs}$ and $L_{model}$ are the luminosity from
observation and model.
"A" is the ratio between the actual mass $m$ in each aperture and the default
value 10$^6$~M$_{\odot}$ provided by the Starburst99 models, $\frac{m}{10^6}$.
$\sigma_{obs}$ accounts for the photometric calibration
and measurement uncertainties. Since we only have four
data points available for each source, we fix both the metallicity and
the mean dust extinction values. Fixing the latter, in particular, overcomes the
age--extinction degeneracy that plagues stellar continuum band fits.
We use as a guidance both the HI map and the
results on HII regions in the M83 outer regions from \citet{gil07}. The HI map
gives mean column densities 3.5$\times10^{20}~cm^{-2}$ and
5$\times10^{20}~cm^{-2}$ for OuterI and OuterL, respectively,
corresponding E(B$-$V)=0.07 and =0.1~\citep{boh78}. As an extreme case, we also
perform fits using E(B$-$V)=0.3, which is at the high--end of the extinction values observed
 by \citet{gil07} in these outer regions.
High extinction values are likely to be applicable to the youngest
SF regions, of which the HII regions of \citet{gil07} are part,
while the lower extinction values are likely to be applicable to
older regions. $A_{FUV}/E(B-V)=8.376$ and $A_{NUV}/E(B-V)=8.741$
for the MW type dust have been used for calculating the
extinction. $L_{3.6}$ and $L_{4.5}$ are assumed to be dust free.
The age and mass of each source corresponding to the minimized
$\chi^{2}$ are listed in table~\ref{t:age}, for both the low and
high extinction values. The best fit values are listed together with the
uncertainty range corresponding to the 90\% confidence level for the
best fits; this is derived from requiring that, in the age and mass parameter space,
$\delta\chi^2$= 4.61 (90\% confidence level for 2 parameters).
Uncertainties are relatively small for our best fit parameters, owing to the
leverage provided by the long--wavelength baseline of the data.

\begin{figure*}[!thb]
  \centerline{
       \epsfig{figure=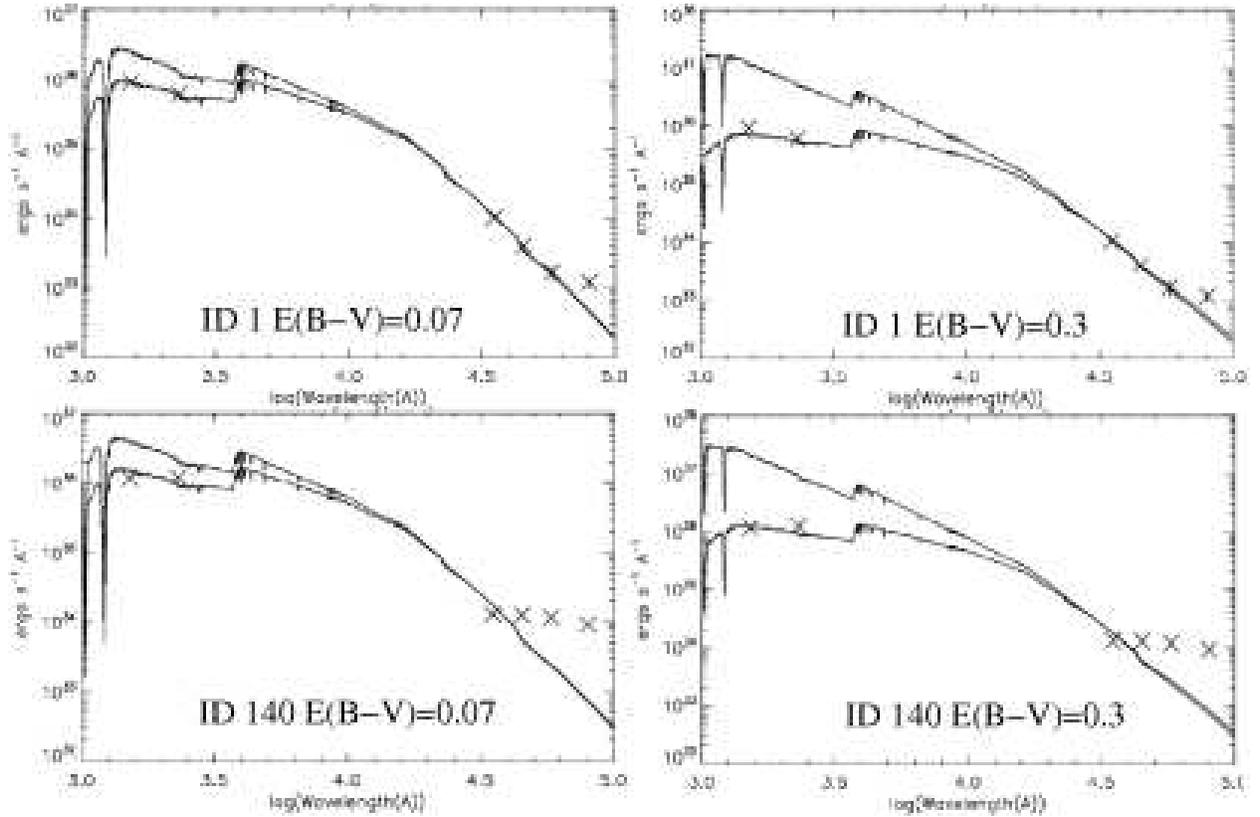,width=1.\textwidth,angle=0}
    }
 \caption{Two examples of our SED fitting results. The top two panels
 show an acceptable fit between data and the models, for the two values of the extinction investigated
 in this work: a low value (here E(B$-$V=0.07 for OuterI) and a high value (E(B$-$V=0.3). An
 `acceptable fit' is defined as a match between the models and the data points to within the 3~$\sigma$
 error bar of each datapoint in the four fitting bands (FUV, NUV, 3.6~$\mu$m and 4.5~$\mu$m).
 In this specific case, we also observe, as in many other cases, a good match also between the
 data and the models at 5.8~$\mu$m, while the 8~$\mu$m emission shows a clear excess due to
 dust emission. The  bottom two panels show an
 example of a source with unusual high excess flux at 4.5, 5.8 and 8.0 $\mu$ (one of the last 16 sources
 in Table~\ref{t:sou}). The two sets of lines are the Starburst99 SEDs without (higher UV values) and
 with (lower UV values) dust extinction.}
 \label{f:compare}
 \end{figure*}

\begin{figure*}[!thb]
  \centerline{
       \epsfig{figure=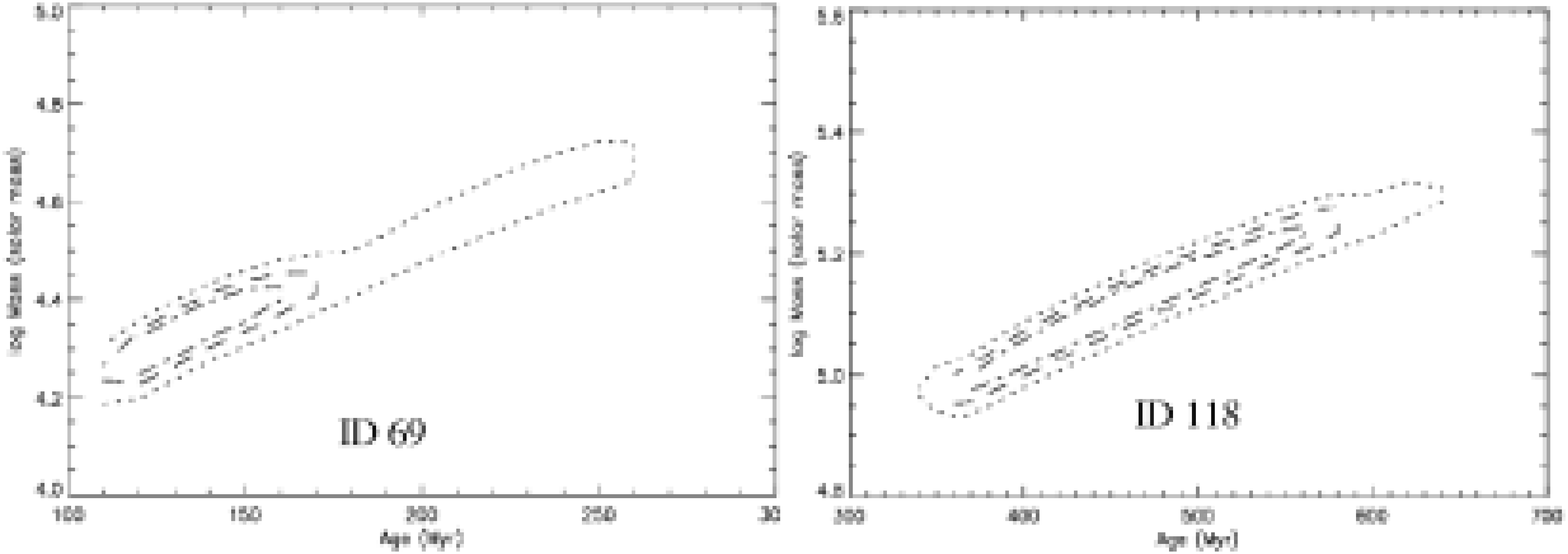,width=1.\textwidth,angle=0}
    }
 \caption{Examples of uncertainty diagrams in mass and age for  two of our regions, one in OuterI
 (ID69) and the other in OuterL (ID118). The plots are for the  low extinction case.
 The lines marking  increasing areas show the 68\%, 90\%, and 99\% confidence levels around the
 best fit values. }
 \label{f:banana}
 \end{figure*}

Fig.~\ref{f:compare} shows two examples of the fitting
results for each extinction value. The two upper panels report the
case of a source with a good match between data and models in the
four primary bands (where a `good match' is defined as agreement
between data and models within the 3~$\sigma$ error bar of each
data point), while the bottom two panels show the case of a source
with unusual high excess 4.5, 5.8 and 8.0 $\mu$m fluxes (one of
the last 16 sources in Table~\ref{t:sou}).  The 68\%, 90\%, and 99\%
confidence levels  on each best fit (age,mass) values are shown in
Fig.~\ref{f:banana} for two representative regions in our sample. These
are both intermediate--luminosity (at 3.6~$\mu$m) sources (see Table~\ref{t:sou}),
one in OuterI and the other in OuterL. The covariance between the two parameters
is such that higher masses correspond to higher ages. This is readily understood
by recalling that the mass is mainly determined by the value of the 3.6~$\mu$m
emission, while the age is constrained by the relative intensity of the UV and
infrared data; increasing the mass by moving the normalization of the SED
at higher 3.6~$\mu$m values will then result in a `flatter' (i.e., older age) best
fitting SED to the UV data.

\begin{deluxetable}{cc|cc|c}
\tabletypesize{\tiny} \tablecolumns{7} \tablecaption{Derived Ages
and Masses for the Sources} \tablewidth{0pt} \tablehead{
   \colhead{} &
   \multicolumn{2}{c}{Age (Myr)} &
   \multicolumn{2}{c}{log Mass ($M_{\bigodot}$)} \\
   \colhead{ID} &
   \colhead{E(B-V)=0.07,0.1\tablenotemark{a}} &
   \colhead{E(B-V)=0.3\tablenotemark{b}}&
   \colhead{E(B-V)=0.07,0.1\tablenotemark{a}} &
   \colhead{E(B-V)=0.3\tablenotemark{b}} \\
   }
\startdata
  1&$260.0^{+10.0}_{-10.0}$&$80.0^{+10.0}_{-10.0}$&$5.67^{+0.01}_{-0.01}$&$5.63^{+0.01}_{-0.01}$\\
  2&$340.0^{+10.0}_{-10.0}$&$100.0^{+10.0}_{-10.0}$&$5.55^{+0.01}_{-0.01}$&$5.48^{+0.01}_{-0.01}$\\
  3&$70.0^{+10.0}_{-10.0}$&$6.0^{+1.0}_{-1.0}$&$5.36^{+0.00}_{-0.00}$&$4.68^{+0.00}_{-0.00}$\\
  4&$250.0^{+10.0}_{-90.0}$&$70.0^{+10.0}_{-10.0}$&$5.12^{+0.02}_{-0.27}$&$5.05^{+0.02}_{-0.08}$\\
  5&$370.0^{+40.0}_{-20.0}$&$110.0^{+10.0}_{-10.0}$&$5.16^{+0.07}_{-0.03}$&$5.00^{+0.02}_{-0.02}$\\
  6&$110.0^{+10.0}_{-10.0}$&$20.0^{+5.0}_{-5.0}$&$5.28^{+0.01}_{-0.01}$&$5.06^{+0.01}_{-0.01}$\\
  7&$640.1^{+20.0}_{-20.0}$&$140.0^{+10.0}_{-10.0}$&$6.39^{+0.00}_{-0.00}$&$5.96^{+0.00}_{-0.00}$\\
  8&$370.0^{+10.0}_{-10.0}$&$110.0^{+10.0}_{-10.0}$&$5.67^{+0.02}_{-0.01}$&$5.51^{+0.01}_{-0.01}$\\
  9&$310.0^{+10.0}_{-10.0}$&$90.0^{+10.0}_{-10.0}$&$5.28^{+0.01}_{-0.02}$&$5.24^{+0.02}_{-0.02}$\\
 10&$900.6^{+20.0}_{-20.0}$&$250.0^{+10.0}_{-10.0}$&$6.26^{+0.00}_{-0.00}$&$5.93^{+0.00}_{-0.00}$\\
 11&$1100.5^{+100.0}_{-100.0}$&$330.0^{+10.0}_{-10.0}$&$6.31^{+0.00}_{-0.00}$&$6.02^{+0.01}_{-0.00}$\\
 12&$640.1^{+40.0}_{-40.0}$&$120.0^{+20.0}_{-10.0}$&$5.69^{+0.04}_{-0.04}$&$5.21^{+0.05}_{-0.01}$\\
 13&$340.0^{+10.0}_{-10.0}$&$100.0^{+10.0}_{-10.0}$&$5.27^{+0.02}_{-0.02}$&$5.19^{+0.02}_{-0.02}$\\
 14&$600.1^{+20.0}_{-20.0}$&$120.0^{+10.0}_{-10.0}$&$5.80^{+0.01}_{-0.01}$&$5.35^{+0.01}_{-0.01}$\\
 15&$350.0^{+10.0}_{-10.0}$&$100.0^{+10.0}_{-10.0}$&$5.71^{+0.01}_{-0.01}$&$5.64^{+0.01}_{-0.01}$\\
 16&$560.3^{+40.0}_{-40.0}$&$110.0^{+10.0}_{-10.0}$&$5.53^{+0.03}_{-0.05}$&$5.13^{+0.01}_{-0.03}$\\
 17&$1200.2^{+101.0}_{-100.0}$&$600.1^{+20.0}_{-20.0}$&$6.47^{+0.00}_{-0.00}$&$6.40^{+0.00}_{-0.00}$\\
 18&$340.0^{+10.0}_{-10.0}$&$100.0^{+10.0}_{-10.0}$&$5.25^{+0.01}_{-0.01}$&$5.18^{+0.01}_{-0.01}$\\
 19&$780.1^{+60.0}_{-20.0}$&$250.0^{+10.0}_{-10.0}$&$6.05^{+0.04}_{-0.01}$&$5.78^{+0.01}_{-0.03}$\\
 20&$490.0^{+50.0}_{-10.0}$&$110.0^{+10.0}_{-10.0}$&$5.52^{+0.06}_{-0.01}$&$5.20^{+0.01}_{-0.01}$\\
 21&$340.0^{+10.0}_{-10.0}$&$100.0^{+10.0}_{-10.0}$&$5.14^{+0.02}_{-0.02}$&$5.06^{+0.02}_{-0.02}$\\
 22&$290.0^{+10.0}_{-10.0}$&$90.0^{+10.0}_{-10.0}$&$5.08^{+0.02}_{-0.03}$&$5.03^{+0.02}_{-0.03}$\\
 23&$1600.7^{+100.0}_{-100.0}$&$900.6^{+20.0}_{-20.0}$&$6.56^{+0.00}_{-0.00}$&$6.73^{+0.00}_{-0.01}$\\
 24&$90.0^{+10.0}_{-20.0}$&$15.0^{+5.0}_{-9.0}$&$4.43^{+0.03}_{-0.15}$&$4.12^{+0.03}_{-0.53}$\\
 25&$6.0^{+1.0}_{-1.0}$&$6.0^{+1.0}_{-1.0}$&$3.39^{+0.02}_{-0.02}$&$4.05^{+0.02}_{-0.02}$\\
 26&$300.0^{+10.0}_{-10.0}$&$35.0^{+5.0}_{-5.0}$&$5.06^{+0.03}_{-0.03}$&$4.59^{+0.02}_{-0.03}$\\
 27&$1200.2^{+101.0}_{-100.0}$&$330.0^{+10.0}_{-10.0}$&$6.14^{+0.01}_{-0.01}$&$5.86^{+0.01}_{-0.01}$\\
 28&$110.0^{+10.0}_{-75.0}$&$0.7^{+1.0}_{-0.6}$&$4.46^{+0.03}_{-0.56}$&$3.44^{+0.06}_{-0.10}$\\
 29&$6.0^{+9.0}_{-1.0}$&$6.0^{+1.0}_{-1.0}$&$3.10^{+0.52}_{-0.11}$&$3.77^{+0.02}_{-0.02}$\\
 30&$3.0^{+2.0}_{-1.0}$&$5.0^{+1.0}_{-1.0}$&$2.77^{+0.19}_{-0.02}$&$3.62^{+0.12}_{-0.02}$\\
 31&$6.0^{+1.0}_{-1.0}$&$6.0^{+1.0}_{-1.0}$&$3.30^{+0.01}_{-0.01}$&$3.97^{+0.01}_{-0.01}$\\
 32&$6.0^{+14.0}_{-1.0}$&$6.0^{+1.0}_{-1.0}$&$2.87^{+0.70}_{-0.11}$&$3.54^{+0.03}_{-0.03}$\\
 33&$20.0^{+5.0}_{-5.0}$&$6.0^{+1.0}_{-1.0}$&$4.65^{+0.00}_{-0.00}$&$4.63^{+0.00}_{-0.00}$\\
 34&$15.0^{+5.0}_{-11.0}$&$6.0^{+1.0}_{-1.0}$&$3.67^{+0.02}_{-0.71}$&$3.82^{+0.02}_{-0.02}$\\
 35&$3.0^{+2.0}_{-1.0}$&$5.0^{+1.0}_{-1.0}$&$2.84^{+0.20}_{-0.03}$&$3.70^{+0.13}_{-0.02}$\\
 36&$90.0^{+10.0}_{-55.0}$&$15.0^{+5.0}_{-8.0}$&$4.56^{+0.02}_{-0.48}$&$4.26^{+0.02}_{-0.41}$\\
 37&$35.0^{+65.0}_{-5.0}$&$3.0^{+1.0}_{-3.0}$&$3.85^{+0.52}_{-0.10}$&$3.21^{+0.17}_{-0.04}$\\
 38&$35.0^{+5.0}_{-5.0}$&$7.0^{+1.0}_{-1.0}$&$4.42^{+0.07}_{-0.02}$&$4.20^{+0.02}_{-0.02}$\\
 39&$100.0^{+10.0}_{-10.0}$&$20.0^{+5.0}_{-5.0}$&$4.52^{+0.03}_{-0.03}$&$4.30^{+0.03}_{-0.03}$\\
 40&$120.0^{+10.0}_{-10.0}$&$25.0^{+5.0}_{-5.0}$&$4.59^{+0.05}_{-0.04}$&$4.43^{+0.03}_{-0.03}$\\
 41&$150.0^{+20.0}_{-30.0}$&$30.0^{+20.0}_{-5.0}$&$4.76^{+0.06}_{-0.12}$&$4.57^{+0.27}_{-0.10}$\\
 42&$310.0^{+10.0}_{-10.0}$&$90.0^{+10.0}_{-10.0}$&$5.16^{+0.02}_{-0.02}$&$5.12^{+0.02}_{-0.02}$\\
 43&$100.0^{+10.0}_{-10.0}$&$20.0^{+5.0}_{-5.0}$&$5.11^{+0.01}_{-0.01}$&$4.90^{+0.01}_{-0.01}$\\
 44&$6.0^{+1.0}_{-1.0}$&$6.0^{+1.0}_{-1.0}$&$3.39^{+0.01}_{-0.01}$&$4.05^{+0.02}_{-0.02}$\\
 45&$280.0^{+20.0}_{-100.0}$&$90.0^{+10.0}_{-10.0}$&$4.95^{+0.03}_{-0.27}$&$4.92^{+0.03}_{-0.05}$\\
 46&$35.0^{+10.0}_{-5.0}$&$15.0^{+5.0}_{-8.0}$&$4.00^{+0.13}_{-0.09}$&$4.18^{+0.03}_{-0.42}$\\
 47&$340.0^{+10.0}_{-10.0}$&$100.0^{+10.0}_{-10.0}$&$5.17^{+0.02}_{-0.02}$&$5.09^{+0.02}_{-0.02}$\\
 48&$540.0^{+20.0}_{-70.0}$&$110.0^{+10.0}_{-10.0}$&$5.55^{+0.03}_{-0.09}$&$5.16^{+0.01}_{-0.01}$\\
 49&$270.0^{+10.0}_{-10.0}$&$90.0^{+10.0}_{-10.0}$&$4.81^{+0.03}_{-0.03}$&$4.78^{+0.03}_{-0.04}$\\
 50&$140.0^{+110.0}_{-20.0}$&$60.0^{+10.0}_{-10.0}$&$4.27^{+0.35}_{-0.10}$&$4.43^{+0.11}_{-0.05}$\\
 51&$340.0^{+10.0}_{-10.0}$&$100.0^{+10.0}_{-10.0}$&$5.31^{+0.01}_{-0.01}$&$5.23^{+0.01}_{-0.01}$\\
 52&$1000.1^{+100.0}_{-20.0}$&$290.0^{+20.0}_{-10.0}$&$5.88^{+0.01}_{-0.01}$&$5.49^{+0.01}_{-0.02}$\\
 53&$110.0^{+10.0}_{-10.0}$&$20.0^{+5.0}_{-5.0}$&$4.45^{+0.02}_{-0.03}$&$4.20^{+0.03}_{-0.03}$\\
 54&$80.0^{+10.0}_{-10.0}$&$6.0^{+1.0}_{-1.0}$&$4.12^{+0.09}_{-0.09}$&$3.36^{+0.03}_{-0.04}$\\
 55&$80.0^{+10.0}_{-35.0}$&$6.0^{+9.0}_{-2.0}$&$4.10^{+0.11}_{-0.31}$&$3.35^{+0.54}_{-0.20}$\\
 56&$120.0^{+10.0}_{-10.0}$&$0.1^{+0.5}_{0.0}$&$4.96^{+0.02}_{-0.02}$&$3.94^{+0.02}_{-0.03}$\\
 57&$110.0^{+20.0}_{-10.0}$&$20.0^{+5.0}_{-10.0}$&$4.28^{+0.09}_{-0.06}$&$4.05^{+0.13}_{-0.41}$\\
 58&$1300.8^{+99.0}_{-101.0}$&$350.0^{+20.0}_{-10.0}$&$6.05^{+0.01}_{-0.01}$&$5.77^{+0.01}_{-0.01}$\\
 59&$300.0^{+20.0}_{-10.0}$&$90.0^{+10.0}_{-10.0}$&$5.05^{+0.03}_{-0.03}$&$5.01^{+0.03}_{-0.03}$\\
 60&$270.0^{+30.0}_{-90.0}$&$80.0^{+10.0}_{-10.0}$&$4.69^{+0.05}_{-0.26}$&$4.65^{+0.06}_{-0.08}$\\
 61&$110.0^{+10.0}_{-10.0}$&$20.0^{+5.0}_{-5.0}$&$4.65^{+0.03}_{-0.03}$&$4.42^{+0.03}_{-0.03}$\\
 62&$35.0^{+10.0}_{-5.0}$&$9.0^{+6.0}_{-6.0}$&$3.77^{+0.13}_{-0.11}$&$3.68^{+0.31}_{-0.55}$\\
 63&$320.0^{+10.0}_{-10.0}$&$100.0^{+10.0}_{-10.0}$&$5.10^{+0.02}_{-0.02}$&$5.03^{+0.04}_{-0.02}$\\
 64&$170.0^{+110.0}_{-20.0}$&$70.0^{+10.0}_{-30.0}$&$4.54^{+0.30}_{-0.07}$&$4.71^{+0.09}_{-0.32}$\\
 65&$430.0^{+90.0}_{-70.0}$&$110.0^{+10.0}_{-10.0}$&$5.13^{+0.12}_{-0.11}$&$4.89^{+0.03}_{-0.03}$\\
 66&$110.0^{+10.0}_{-10.0}$&$20.0^{+5.0}_{-20.0}$&$4.29^{+0.05}_{-0.07}$&$4.07^{+0.05}_{-0.92}$\\
 67&$250.0^{+10.0}_{-30.0}$&$70.0^{+10.0}_{-10.0}$&$4.97^{+0.02}_{-0.09}$&$4.88^{+0.02}_{-0.03}$\\
 68&$250.0^{+10.0}_{-100.0}$&$70.0^{+10.0}_{-20.0}$&$4.72^{+0.04}_{-0.32}$&$4.64^{+0.09}_{-0.18}$\\
 69&$130.0^{+40.0}_{-20.0}$&$25.0^{+35.0}_{-5.0}$&$4.32^{+0.13}_{-0.10}$&$4.13^{+0.44}_{-0.06}$\\
 70&$120.0^{+160.0}_{-120.0}$&$0.3^{+25.0}_{-0.2}$&$3.90^{+0.44}_{-1.96}$&$2.91^{+0.82}_{-0.28}$\\
 71&$110.0^{+10.0}_{-10.0}$&$20.0^{+5.0}_{-5.0}$&$4.40^{+0.04}_{-0.04}$&$4.17^{+0.11}_{-0.04}$\\
 72&$280.0^{+10.0}_{-10.0}$&$90.0^{+10.0}_{-10.0}$&$4.89^{+0.03}_{-0.03}$&$4.86^{+0.03}_{-0.03}$\\
 73&$540.0^{+40.0}_{-90.0}$&$110.0^{+10.0}_{-10.0}$&$5.23^{+0.05}_{-0.11}$&$4.84^{+0.02}_{-0.02}$\\
 74&$450.0^{+110.0}_{-70.0}$&$110.0^{+10.0}_{-10.0}$&$5.06^{+0.14}_{-0.10}$&$4.77^{+0.02}_{-0.02}$\\
 75&$540.0^{+40.0}_{-50.0}$&$110.0^{+10.0}_{-10.0}$&$5.36^{+0.04}_{-0.06}$&$4.97^{+0.01}_{-0.02}$\\
 76&$270.0^{+20.0}_{-90.0}$&$80.0^{+10.0}_{-10.0}$&$4.94^{+0.03}_{-0.25}$&$4.90^{+0.03}_{-0.03}$\\
 77&$1100.5^{+100.0}_{-100.0}$&$340.0^{+10.0}_{-10.0}$&$6.03^{+0.01}_{-0.01}$&$5.73^{+0.01}_{-0.01}$\\
 78&$270.0^{+10.0}_{-10.0}$&$90.0^{+10.0}_{-10.0}$&$4.96^{+0.03}_{-0.04}$&$4.93^{+0.03}_{-0.04}$\\
 79&$350.0^{+30.0}_{-10.0}$&$100.0^{+10.0}_{-10.0}$&$5.03^{+0.02}_{-0.03}$&$4.96^{+0.02}_{-0.03}$\\
 80&$370.0^{+60.0}_{-20.0}$&$100.0^{+10.0}_{-10.0}$&$5.00^{+0.09}_{-0.03}$&$4.93^{+0.02}_{-0.12}$\\
 81&$450.0^{+90.0}_{-80.0}$&$110.0^{+10.0}_{-10.0}$&$5.12^{+0.11}_{-0.13}$&$4.84^{+0.02}_{-0.02}$\\
 82&$420.0^{+70.0}_{-50.0}$&$110.0^{+10.0}_{-10.0}$&$5.20^{+0.09}_{-0.08}$&$4.96^{+0.02}_{-0.02}$\\
 83&$580.0^{+40.0}_{-60.0}$&$120.0^{+10.0}_{-10.0}$&$5.42^{+0.03}_{-0.07}$&$4.98^{+0.03}_{-0.02}$\\
 84&$280.0^{+20.0}_{-110.0}$&$80.0^{+10.0}_{-45.0}$&$4.73^{+0.04}_{-0.29}$&$4.69^{+0.05}_{-0.45}$\\
 85(XUV1)\tablenotemark{c}&$80.0^{+10.0}_{-10.0}$&$6.0^{+1.0}_{-1.0}$&$4.93^{+0.01}_{-0.08}$&$4.30^{+0.01}_{-0.01}$\\
 86(XUV3)\tablenotemark{c}&$120.0^{+10.0}_{-10.0}$&$20.0^{+5.0}_{-5.0}$&$5.36^{+0.01}_{-0.01}$&$5.21^{+0.01}_{-0.01}$\\
 87&$100.0^{+10.0}_{-10.0}$&$15.0^{+5.0}_{-8.0}$&$4.63^{+0.03}_{-0.03}$&$4.42^{+0.03}_{-0.42}$\\
 88&$290.0^{+10.0}_{-110.0}$&$60.0^{+10.0}_{-10.0}$&$5.00^{+0.03}_{-0.28}$&$4.90^{+0.10}_{-0.11}$\\
 89&$420.0^{+50.0}_{-20.0}$&$100.0^{+10.0}_{-10.0}$&$5.61^{+0.07}_{-0.04}$&$5.49^{+0.01}_{-0.01}$\\
 90(XUV2)\tablenotemark{c}&$90.0^{+10.0}_{-10.0}$&$6.0^{+1.0}_{-1.0}$&$4.68^{+0.06}_{-0.02}$&$3.99^{+0.02}_{-0.02}$\\
 91&$600.1^{+40.0}_{-20.0}$&$110.0^{+10.0}_{-10.0}$&$5.91^{+0.04}_{-0.01}$&$5.50^{+0.01}_{-0.01}$\\
 92&$780.1^{+60.0}_{-80.0}$&$110.0^{+10.0}_{-10.0}$&$5.89^{+0.05}_{-0.06}$&$5.35^{+0.01}_{-0.02}$\\
 93&$1000.1^{+100.0}_{-60.0}$&$140.0^{+10.0}_{-10.0}$&$6.23^{+0.01}_{-0.03}$&$5.61^{+0.01}_{-0.01}$\\
 94&$90.0^{+10.0}_{-30.0}$&$6.0^{+1.0}_{-1.0}$&$4.40^{+0.05}_{-0.22}$&$3.72^{+0.03}_{-0.03}$\\
 95&$80.0^{+10.0}_{-40.0}$&$6.0^{+1.0}_{-1.0}$&$4.52^{+0.08}_{-0.34}$&$3.90^{+0.02}_{-0.03}$\\
 96&$100.0^{+10.0}_{-10.0}$&$7.0^{+8.0}_{-3.0}$&$4.74^{+0.02}_{-0.03}$&$4.13^{+0.41}_{-0.33}$\\
 97(XUV4)\tablenotemark{c}&$110.0^{+10.0}_{-10.0}$&$15.0^{+5.0}_{-5.0}$&$5.24^{+0.01}_{-0.01}$&$5.00^{+0.01}_{-0.01}$\\
 98&$100.0^{+10.0}_{-10.0}$&$7.0^{+8.0}_{-1.0}$&$4.72^{+0.02}_{-0.02}$&$4.11^{+0.41}_{-0.02}$\\
 99&$180.0^{+10.0}_{-10.0}$&$25.0^{+5.0}_{-5.0}$&$5.01^{+0.01}_{-0.01}$&$4.77^{+0.01}_{-0.01}$\\
100&$90.0^{+10.0}_{-20.0}$&$6.0^{+1.0}_{-1.0}$&$4.48^{+0.02}_{-0.14}$&$3.80^{+0.02}_{-0.02}$\\
101&$90.0^{+10.0}_{-10.0}$&$6.0^{+1.0}_{-1.0}$&$4.65^{+0.02}_{-0.07}$&$3.97^{+0.02}_{-0.02}$\\
102&$60.0^{+30.0}_{-54.0}$&$6.0^{+1.0}_{-1.0}$&$4.27^{+0.21}_{-1.27}$&$3.81^{+0.03}_{-0.12}$\\
103&$30.0^{+50.0}_{-30.0}$&$6.0^{+1.0}_{-1.0}$&$3.92^{+0.49}_{-1.10}$&$3.78^{+0.03}_{-0.13}$\\
104&$310.0^{+30.0}_{-140.0}$&$30.0^{+40.0}_{-5.0}$&$4.82^{+0.06}_{-0.33}$&$4.38^{+0.41}_{-0.10}$\\
105&$70.0^{+10.0}_{-10.0}$&$6.0^{+1.0}_{-1.0}$&$4.52^{+0.09}_{-0.11}$&$3.97^{+0.03}_{-0.03}$\\
106&$100.0^{+10.0}_{-10.0}$&$6.0^{+9.0}_{-1.0}$&$4.72^{+0.02}_{-0.05}$&$3.97^{+0.51}_{-0.02}$\\
107&$280.0^{+10.0}_{-10.0}$&$70.0^{+10.0}_{-10.0}$&$5.48^{+0.01}_{-0.01}$&$5.45^{+0.01}_{-0.01}$\\
108&$120.0^{+20.0}_{-10.0}$&$20.0^{+5.0}_{-5.0}$&$4.76^{+0.10}_{-0.02}$&$4.60^{+0.02}_{-0.02}$\\
109&$80.0^{+20.0}_{-70.0}$&$6.0^{+9.0}_{-1.0}$&$4.11^{+0.14}_{-1.07}$&$3.49^{+0.53}_{-0.15}$\\
110&$290.0^{+30.0}_{-130.0}$&$30.0^{+40.0}_{-5.0}$&$4.72^{+0.06}_{-0.35}$&$4.29^{+0.40}_{-0.12}$\\
111&$340.0^{+10.0}_{-10.0}$&$90.0^{+10.0}_{-10.0}$&$5.30^{+0.02}_{-0.02}$&$5.30^{+0.02}_{-0.02}$\\
112&$35.0^{+65.0}_{-35.0}$&$4.0^{+11.0}_{-2.0}$&$3.70^{+0.57}_{-1.28}$&$3.29^{+0.73}_{-0.18}$\\
113&$250.0^{+10.0}_{-50.0}$&$60.0^{+10.0}_{-35.0}$&$5.32^{+0.03}_{-0.15}$&$5.26^{+0.03}_{-0.43}$\\
114&$6.0^{+74.0}_{-6.0}$&$6.0^{+1.0}_{-1.0}$&$2.69^{+1.42}_{-0.32}$&$3.48^{+0.04}_{-0.13}$\\
115&$180.0^{+120.0}_{-30.0}$&$25.0^{+45.0}_{-5.0}$&$4.51^{+0.31}_{-0.12}$&$4.27^{+0.44}_{-0.06}$\\
116&$80.0^{+10.0}_{-35.0}$&$6.0^{+1.0}_{-1.0}$&$4.47^{+0.09}_{-0.30}$&$3.85^{+0.04}_{-0.12}$\\
117&$6.0^{+1.0}_{-1.0}$&$6.0^{+1.0}_{-1.0}$&$3.05^{+0.02}_{-0.02}$&$3.80^{+0.02}_{-0.02}$\\
118&$450.0^{+130.0}_{-90.0}$&$100.0^{+10.0}_{-10.0}$&$5.11^{+0.17}_{-0.15}$&$4.95^{+0.04}_{-0.04}$\\
119&$6.0^{+74.0}_{-6.0}$&$6.0^{+1.0}_{-1.0}$&$2.81^{+1.44}_{-0.32}$&$3.60^{+0.06}_{-0.17}$\\
120&$0.3^{+35.0}_{-0.2}$&$5.0^{+1.0}_{-1.0}$&$2.89^{+1.10}_{-0.22}$&$3.67^{+0.14}_{-0.11}$\\
121&$100.0^{+10.0}_{-10.0}$&$15.0^{+5.0}_{-8.0}$&$4.33^{+0.06}_{-0.08}$&$4.10^{+0.05}_{-0.43}$\\
122&$1400.1^{+101.0}_{-99.0}$&$1100.5^{+100.0}_{-100.0}$&$6.00^{+0.01}_{-0.01}$&$6.05^{+0.01}_{-0.01}$\\
123&$700.0^{+60.0}_{-100.0}$&$110.0^{+10.0}_{-10.0}$&$5.47^{+0.06}_{-0.10}$&$4.98^{+0.02}_{-0.03}$\\
124&$0.3^{+50.0}_{-0.2}$&$5.0^{+10.0}_{-2.0}$&$2.58^{+1.30}_{-0.25}$&$3.36^{+0.68}_{-0.25}$\\
125&$640.1^{+160.0}_{-80.0}$&$110.0^{+10.0}_{-10.0}$&$5.49^{+0.13}_{-0.07}$&$5.05^{+0.03}_{-0.03}$\\
126&$110.0^{+30.0}_{-85.0}$&$8.0^{+12.0}_{-6.0}$&$4.17^{+0.14}_{-0.76}$&$3.61^{+0.50}_{-0.55}$\\
127&$110.0^{+50.0}_{-10.0}$&$9.0^{+11.0}_{-9.0}$&$4.12^{+0.20}_{-0.08}$&$3.60^{+0.47}_{-0.61}$\\
128&$7.0^{+93.0}_{-7.0}$&$6.0^{+9.0}_{-2.0}$&$2.57^{+1.40}_{-0.47}$&$3.23^{+0.54}_{-0.24}$\\
129&$350.0^{+70.0}_{-20.0}$&$90.0^{+10.0}_{-10.0}$&$4.94^{+0.10}_{-0.04}$&$4.94^{+0.04}_{-0.05}$\\
130&$940.2^{+60.0}_{-60.0}$&$120.0^{+20.0}_{-10.0}$&$5.74^{+0.04}_{-0.04}$&$5.09^{+0.06}_{-0.02}$\\
131&$130.0^{+120.0}_{-20.0}$&$20.0^{+5.0}_{-10.0}$&$4.44^{+0.39}_{-0.12}$&$4.24^{+0.05}_{-0.38}$\\
132&$70.0^{+20.0}_{-10.0}$&$6.0^{+1.0}_{-1.0}$&$4.17^{+0.14}_{-0.11}$&$3.62^{+0.03}_{-0.04}$\\
133&$80.0^{+10.0}_{-20.0}$&$6.0^{+1.0}_{-1.0}$&$4.16^{+0.09}_{-0.17}$&$3.53^{+0.04}_{-0.04}$\\
134&$110.0^{+10.0}_{-10.0}$&$15.0^{+5.0}_{-8.0}$&$4.28^{+0.04}_{-0.07}$&$4.02^{+0.04}_{-0.42}$\\
135&$1100.5^{+100.0}_{-100.0}$&$260.0^{+10.0}_{-10.0}$&$6.06^{+0.01}_{-0.01}$&$5.83^{+0.01}_{-0.01}$\\
136&$140.0^{+110.0}_{-30.0}$&$20.0^{+5.0}_{-10.0}$&$4.32^{+0.36}_{-0.14}$&$4.09^{+0.11}_{-0.40}$\nl
\cline{1-5}
    \multicolumn{5}{c}{} \\
    \multicolumn{5}{c}{The sources with unusual high excess 4.5, 5.8
and 8.0 $\mu$m flux} \nl
\multicolumn{5}{c}{}\\
\cline{1-5}
137&$820.0^{+60.0}_{-40.0}$&$250.0^{+10.0}_{-10.0}$&$6.08^{+0.03}_{-0.03}$&$5.79^{+0.01}_{-0.03}$\\
138&$100.0^{+10.0}_{-10.0}$&$20.0^{+5.0}_{-5.0}$&$4.99^{+0.01}_{-0.01}$&$4.78^{+0.01}_{-0.01}$\\
139&$560.3^{+20.0}_{-20.0}$&$110.0^{+10.0}_{-10.0}$&$6.01^{+0.00}_{-0.01}$&$5.60^{+0.01}_{-0.01}$\\
140&$250.0^{+10.0}_{-10.0}$&$70.0^{+10.0}_{-10.0}$&$5.87^{+0.00}_{-0.00}$&$5.78^{+0.00}_{-0.00}$\\
141&$420.0^{+10.0}_{-20.0}$&$110.0^{+10.0}_{-10.0}$&$5.71^{+0.01}_{-0.03}$&$5.47^{+0.01}_{-0.01}$\\
142&$900.6^{+20.0}_{-20.0}$&$250.0^{+10.0}_{-10.0}$&$6.24^{+0.00}_{-0.01}$&$5.91^{+0.00}_{-0.00}$\\
143&$1100.5^{+100.0}_{-100.0}$&$1100.5^{+100.0}_{-100.0}$&$6.25^{+0.00}_{-0.00}$&$6.25^{+0.00}_{-0.00}$\\
144&$600.1^{+80.0}_{-20.0}$&$120.0^{+10.0}_{-10.0}$&$5.61^{+0.06}_{-0.01}$&$5.16^{+0.01}_{-0.01}$\\
145&$1100.5^{+100.0}_{-100.0}$&$290.0^{+10.0}_{-10.0}$&$6.05^{+0.01}_{-0.01}$&$5.80^{+0.01}_{-0.01}$\\
146&$1000.1^{+100.0}_{-20.0}$&$290.0^{+10.0}_{-10.0}$&$5.89^{+0.01}_{-0.01}$&$5.50^{+0.01}_{-0.01}$\\
147&$1100.5^{+100.0}_{-100.0}$&$250.0^{+10.0}_{-40.0}$&$5.69^{+0.01}_{-0.01}$&$5.46^{+0.01}_{-0.12}$\\
148&$680.4^{+120.0}_{-40.0}$&$140.0^{+10.0}_{-20.0}$&$5.57^{+0.09}_{-0.04}$&$5.11^{+0.01}_{-0.06}$\\
149&$600.1^{+40.0}_{-60.0}$&$120.0^{+10.0}_{-10.0}$&$5.37^{+0.04}_{-0.06}$&$4.92^{+0.02}_{-0.02}$\\
150&$320.0^{+10.0}_{-10.0}$&$80.0^{+10.0}_{-10.0}$&$5.52^{+0.01}_{-0.01}$&$5.50^{+0.03}_{-0.01}$\\
151&$400.0^{+10.0}_{-10.0}$&$100.0^{+10.0}_{-10.0}$&$6.23^{+0.00}_{-0.02}$&$6.14^{+0.00}_{-0.00}$\\
152&$600.1^{+40.0}_{-40.0}$&$110.0^{+10.0}_{-10.0}$&$5.62^{+0.04}_{-0.04}$&$5.21^{+0.02}_{-0.02}$\\
\enddata
\tablenotetext{a}{SED fitting results for the 152 sources in the two OuterI and OuterL fields.
The values of the color excess used in the fits  are E(B$-V$)=0.07 for OuterI and
E(B$-V$)=0.1 for OuterL. These values are derived from the average
HI column densities in each region,  $N_{HI}$=3.5$\times10^{20}~cm^{-2}$ for OuterI and
$N_{HI}$=5$\times10^{20}~cm^{-2}$ for OuterL, and the extinction--to--HI~column~density ratio
of \citet{boh78}.}
 \tablenotetext{b}{SED fitting results, applying a larger value of the extinction, E(B-V)=0.3, as
 derived by \citet{gil07} for a few of the HII regions detected in the M83 outskirts.}
 \tablenotetext{c}{The IDs with added `XUVn' labels give the cross-identification of the sources
 in common with those of \citet{gil07}.}
 \label{t:age}
\end{deluxetable}

\begin{figure*}[!thb]
  \centerline{
       \epsfig{figure=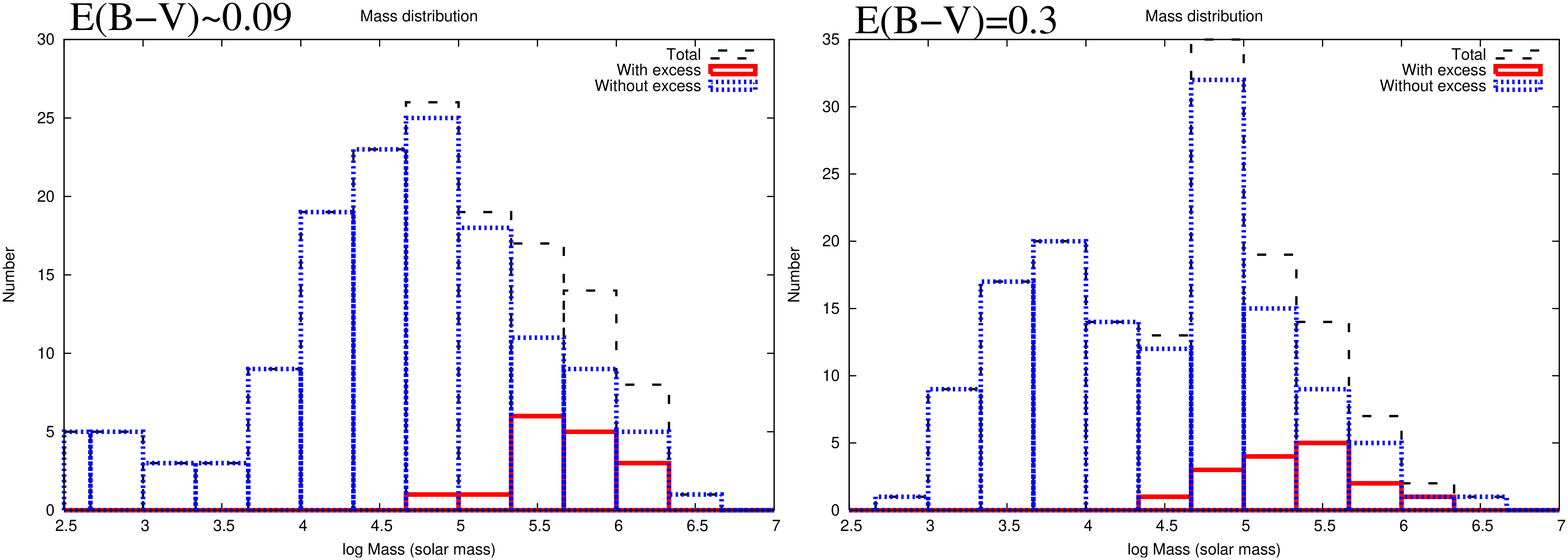,width=1.\textwidth,angle=0}
    }
 \caption{Histogram of the mass distribution of the sample, as obtained from the
 SED fitting,  for both the extinction values adopted: E(B$-$V)$\sim$0.09 (indicative of the
 actual values of E(B$-$V)$\sim$0.07 for OuterI and 0.1 for OuterL, left panel) and
 E(B$-$V)=0.3 (right panel. Black dash line for the total of 152
 sources, blue dot line for the 136 final sources and red line for
 the sources with unusually large 4.5, 5.8 and 8.0 $\mu$m excess.}
 \label{f:mass}
 \end{figure*}

\begin{figure*}[!thb]
  \centerline{
       \epsfig{figure=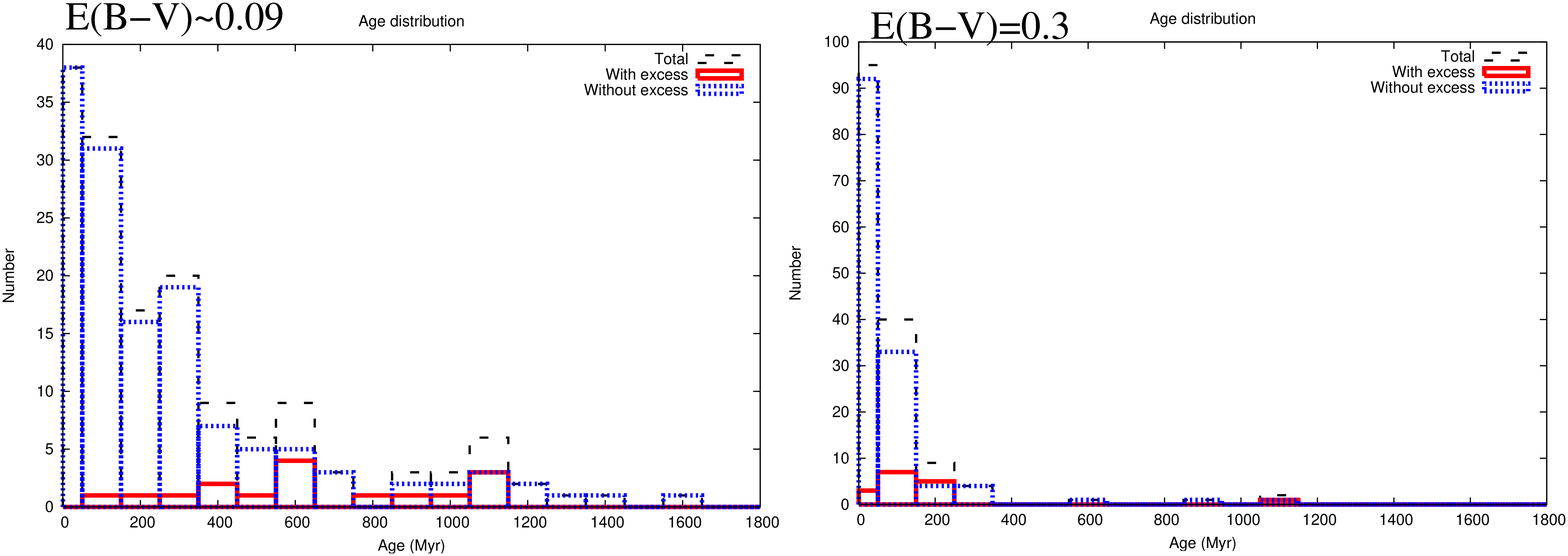,width=1.\textwidth,angle=0}
    }
 \caption{As in Figure~\ref{f:mass}, but for the age distribution of the sources.}
 \label{f:age}
 \end{figure*}

The distributions of the best fit masses and ages for the 152 sources (also separating the 136 without
infrared excess from those with infrared excess, see above) are shown in Fig.~\ref{f:mass}
and Fig.~\ref{f:age}, for both the low and high extinction values. As expected from the
age--extinction degeneracy, the fits with the higher extinction value produce
significantly lower age values, as these are mostly determined by the ratio of UV to IR flux.
However, the mean mass, which is determined mainly by the 3.6 $\mu$m flux value,
is relatively well constrained (to better than a factor of 2).
For the low extinction values, the mass distribution has a peak around
$10^{4.7}~M_{\bigodot}$, with a median value around
$10^{4.9}~M_{\bigodot}$. For high extinction (E(B-V)=0.3), the median value
decreases to $10^{4.7}~M_{\bigodot}$, i.e. by about 60\%. Masses as large as
$10^{6}~M_{\bigodot}$ are also found in our sample. In Fig.3 of
\citet{thi05}, the masses of several UV bright regions are
larger than $10^{5}~M_{\bigodot}$, but their derived masses tend to be on average
slightly smaller than the ones we derive here, possibly owing to the better constraint offered
by the \irac\ measurements.

\begin{figure*}[!thb]
  \centerline{
       \epsfig{figure=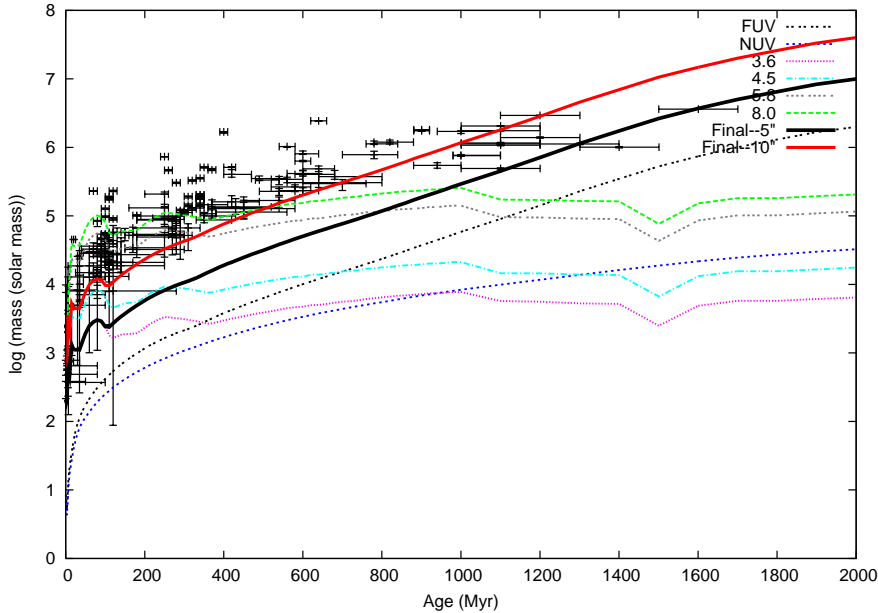,width=0.5\textwidth,angle=-90}
    }
 \caption{Mass (in logarithm scale and in units of M$_{\odot}$) as a function of age (in
 linear scale and in units of Myr) for all our 152 sources.  The thin lines identified with
 a bandpass name show the 1~$\sigma$ detection limit in that bandpass for a region of
 5$^{\prime\prime}$ radius for the six  \galex\ and \spitzer\  filters. The actual detection limit
 for our sources is a complex combination of selection criteria involving both UV and infrared
 detections, as described in section~\ref{s:sample}.  In particular, the 3.6~$\mu$m detection limit
 is the dominant  selection criterion at young ages, while the condition that FUV$>$5~$\sigma$
 dominates at older ages. The 1~$\sigma$ detection limit from those
 combined criteria is shown in the Figure as a thick black line (Final-5") for 5$^{\prime\prime}$ radius
 apertures and a thick red line (Final-10") for 10$^{\prime\prime}$ radius  apertures.
For sources younger than $\approx$50--100~Myr, the dominant selection criterion is the 3.6~$\mu$m
detection, while for older ages limits in the FUV detections dominate the selection.
}
 \label{f:agevsmass}
 \end{figure*}

Star formation has been an ongoing process in the outskirts of M83 for at
least the past 1~Gyr (Fig.~\ref{f:age}). There is a noticeable increase in the number of sources towards
younger ages,  and the median age for the sample is indeed $\sim$180~Myr; however this increase
could be due selection effects. Our UV--based source selection method gives preference to
young stellar population, and we miss most of the old stellar
population due to their faint UV emission. This effect can be seen
clearly from Fig.~\ref{f:agevsmass}.  In this Figure,
the distribution of age versus mass  seems to follow
a trend of increasing mean age for increasing mass. While the empty region
in the bottom--right corner of the plot (regions of old age and low total mass) is
due to instrumental detection limits, the lack of young and massive stellar populations
points to possibly another type of selection effect, the so--called `size--of--sample' effect
\citep[e.g.,][and references therein]{hun03}. For constant star formation, the relation between
the maximum cluster mass observed and the age of the clusters is related by:
M$_{MAX}\propto$(age)$^{1/\alpha -1)}$, where $\alpha$ is the exponent of the initial cluster mass
function. By fitting a linear relation to the upper envelope of our data in a Log(Mass)--Log(age) plot,
we derive $\alpha\approx$2, a value consistent with results for cluster mass functions
\citep{hun03}.

Within our sample of 136 sources, 13 ($\sim$10.5\%)  have
ages of less than 10 Myr. This is the age range when ionizing radiation is expected
to be produced by massive stars. Our observed fraction of 10.5\% ionizing sources
is similar to the fraction of UV spots with
H$\alpha$ counterparts reported in~\citet{thi05}. For the higher extinction value (Fig.~\ref{f:age},
right panel), we obtain that 28.3\% sources are younger than 10 Myr,
which is mildly inconsistent with the dearth of H$\alpha$ sources in the outer regions of M83. Thus,
we infer that E(B-V)$\sim$0.1 is a more realistic extinction value for the
majority of the sources in our regions.

Not surprisingly, the 16 regions with excess 4.5, 5.8, and
8.0~$\mu$m fluxes have systematically larger masses and on average older ages
than the rest of the sample, reinforcing the possibility that the IR excess
is due to contamination from unrecognized foreground or background
sources that cannot be separated from the in--situ source of
UV emission. Indeed, including the 2MASS J,H, and K$_s$ photometry in the SED
fitting for these 16 sources shows that the 4.5, 5.8, and 8.0~$\mu$m IRAC fluxes are
systematically larger (in some cases by more than an order of magnitude) than the best
fit through the GALEX, 2MASS and IRAC 3.6~$\mu$m data. Despite the large uncertainties
in the 2MASS datapoints, this result confirms that the last 16 regions of Table~\ref{t:sou}
are contaminated by (likely) background galaxies at large distances.

The inclusion of the 2MASS datapoints in the SED fitting of the 136 `bonafide' sources
supports the results obtained for the GALEX$+$IRAC--only fits:  single--age populations
are still an acceptable fit to the data. The fits that include the 2MASS data produce median
ages only marginally younger than without the 2MASS data (160~Myr versus 180~Myr),
and median masses that are a factor 1.6 smaller ($10^{4.7}~M_{\bigodot}$
versus $10^{4.9}~M_{\bigodot}$). The 2MASS data, thus, provide a sanity check that support our
baseline results.

\begin{figure*}[!thb]
       \plottwo{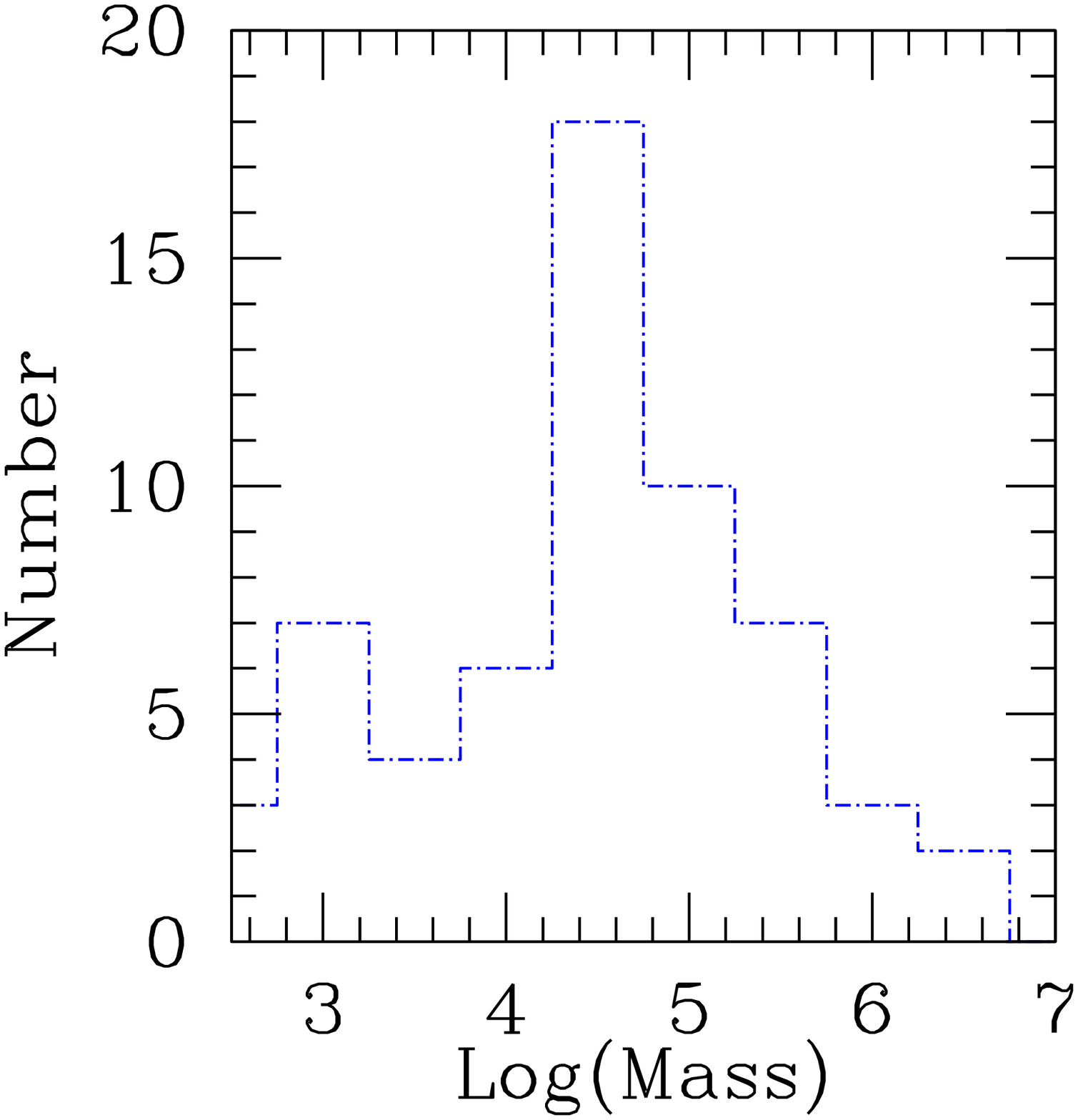}{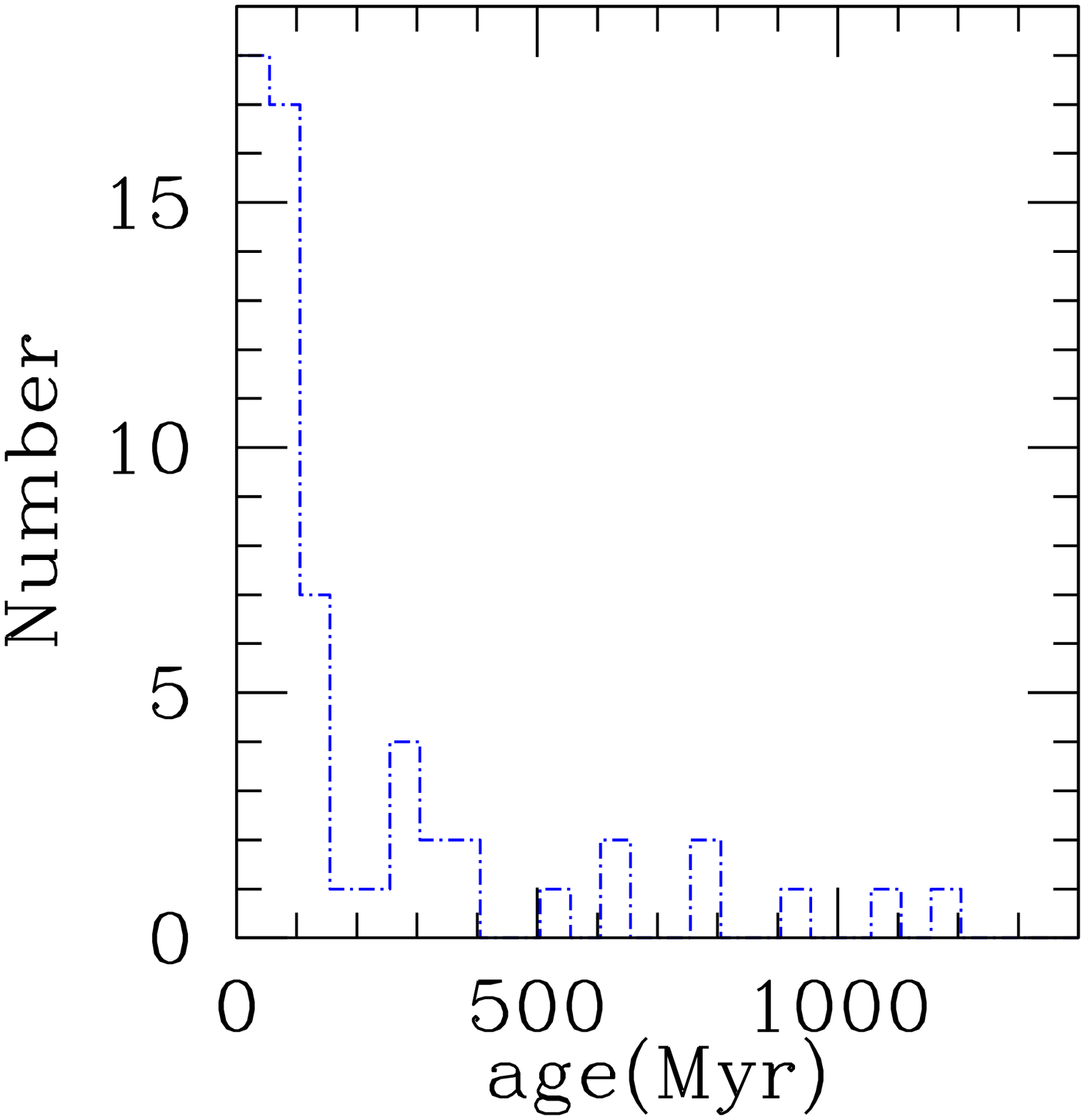}
 \caption{Mass (left, in units of M$_{\odot}$) and age (right, in units of Myr) distributions for the
 60 sources located in regions of high HI column density in both OuterI and OuterL. The low
 extinction values have been adopted in the SED fits.}
 \label{f:60sou}
 \end{figure*}

To further test whether even our bonafide 136 final sources may be contaminated
by unrecognized sources unrelated to M83, we have isolated those sources that are
located in regions of HI column densities above the median of the regions
(N(HI)$>$2.2$\times$10$^{20}$~cm$^{-2}$). The hypothesis behind this selection is that
sources located in correspondence of relatively high HI column densities are more likely
to be physically associated with M83. There are 60 such sources (44\% of the total), and their
mass and age distributions are shown in Figure~\ref{f:60sou}. Both distributions are consistent
with those of the whole sample of 136 sources, with a median value for the age which is about
half that of the whole sample ($\sim$90~Myr
versus 180~Myr), and a factor 2.5 lower median mass (10$^{4.5}$~M$_{\odot}$ versus
10$^{4.9}$~M$_{\odot}$).

\subsection{Comparison with H$\alpha$ observations} \label{s:halpha}
~\citet{gil07} obtained optical spectra in the Southern parts of the
M83 outskirts, and four emission line sources fall within our OuterL field:
XUV01-04 ~\citep[Table 2 of ][]{gil07}. The cross-indentifications with our
sources are given in Table~\ref{t:age}.
We note significant spatial off-sets between XUV3 and XUV4 and their
counterparts in our sample (86 and 97, see Table~\ref{t:age}). However, since
regions 86 and 97 are the brightest and closest UV spots around these two H$\alpha$--emitting
regions, we believe that they come from the same objects.

From Table~\ref{t:age} and Table~\ref{t:XUV}, all regions in common with the sample of
\citet{gil07} have ages around 100~Myr,
which would indicate that they should not be line emitting sources, in
apparent contradiction with the results of  \citet{gil07}. However, if we take as
examples XUV2 and XUV3, the extinction
values measured by those authors for these two regions are 0.23 and 0.29, respectively.
 If we adopt the larger extinction value E(B$-$V)=0.3 in the SED fitting, we obtain that
 XUV2 becomes 6 Myr old and XUV3 20 Myr old. All four sources in common with \citet{gil07}
 have indeed ages equal to or younger than $\sim$20~Myr for the higher extinction value
 (Table~\ref{t:age}), which are not incompatible with those sources being line--emitting ones.
 For XUV3, HST images (Thilker et al, 2007, in
preparation) show the presence of an older (red) stellar population within the source.
In our \irac\ observations, XUV4 is polluted by a nearby foreground
star, which may be partially responsible for an artificial increase of the derived age.
XUV1 has the largest H$\beta$ flux among the four sources in common with
the sample of \citet{gil07} and has a very small extinction as derived from the optical
spectra. \citet{gil07} suggests that the H$\beta$
flux of XUV1 could be produced by a single young massive star
3 Myr old and with a mass of 40 $M_{\bigodot}$. Both the age (80~Myr for the small
extinction value, see Table~\ref{t:XUV}) and the total mass (which depends mainly
from the \irac\ 3.6~$\mu$m flux and implies that the total mass of
stars above  40 $M_{\bigodot}$ is 4300 $M_{\bigodot}$) that we derive for XUV1 are
again in apparent contradiction with these results. We suggest that, for XUV1
 multiple stellar populations, spanning a range of ages, are present within the
region enclosed by our apertures, with the older ones carrying most of the mass and
responsible for the  \irac\ flux and the younger ones responsible for the
strong H$\alpha$ flux. This hypothesis is also in line with the HST observational results
for XUV3.

\begin{deluxetable}{cccccc}
  \tabletypesize{\tiny}
  \tablecaption{The counterparts of the $H\alpha$ sources}
  \tablewidth{0pt}
  \tablehead{
   \colhead{XUV region Name$^a$} &
   \colhead{ID$^b$} &
   \colhead{$f_{H\beta 0}$ (ergs $s^{-1}$ $cm^{-2}$)$^c$} &
   \colhead{distance (arcsecond)} &
   \colhead{Age$^d$ (Myr)} &
   \colhead{log (Mass$^d$ (solar mass))}
   }
  \startdata
  XUV1 & 85 & 1.11$\times10^{-15}$ & 1.35 &80. & 4.9 \\
  XUV2 & 90 & 2.39$\times10^{-16}$ & 4.4 & 90. & 4.7 \\
  XUV3 & 86  & 2.07$\times10^{-16}$ & 12.4 & 120. & 5.4 \\
  XUV4 & 97 & 1.60$\times10^{-16}$ & 8.1 & 110. & 5.2 \\
  \enddata
\tablecomments{(a) The XUV region naming convention from Table~2 of
~\citet{gil07}, (b) The ID in table.~\ref{t:sou}, (c) The H$\beta$
flux after the extinction correction from ~\citet{gil07} and (d)
The age and mass calculated from our SED fitting method, for the low
extinction value E(B$-$V)=0.1, see Table~\ref{t:age}.} \label{t:XUV}
\end{deluxetable}

\subsection{Dust Emission} \label{s:dust}
Observational evidence for extended dust emission around galaxy disks
has increased in recent times. Extended (beyond the optical disk) dust emission
has been observed in a variety of galaxies using both ISO and Spitzer, from
large edge--on spirals like  NGC891 \citep{pop03} to dwarfs like UGC10445 \citep{hin06}.
Non-negligible extinction has been measured up to 2 effective radii in spiral galaxies using
the overlapping pair method
\citep[e.g.,][]{hol07}. \galex\ observations of the edge-on starburst
galaxies M82 and NGC 253 also show unusual high UV luminosity in
the halo, which cannot be explained by shock--heated or photoionized
gas and could have dust--scattering origin ~\citep{hoo05}, in agreement with
evidence for dust emission in these same areas ~\citep{eng06}. Presence
of dust in the outer regions of galaxies can provide crucial insights into the
metal pollution of  the intergalactic medium.

By calculating the excess infrared flux over the SED best fitting models
for our 136 sources, we get that about 14$\pm$6\%, 24$\pm$30\% and
67$\pm$26\% of the total flux at 4.5, 5.8 and 8.0 $\mu m$, respectively, results
from the contribution of a non-stellar (dust emission) component.

The 8~$\mu$m dust--only emission can be used to derive an approximate value of the
SFR in the two regions, keeping in mind a number of caveats about using
the PAH emission for tracing star formation~\citep{cal07}: it is very sensitive
to both metallicity \citep{eng05} and the star formation history of the region
under consideration. In particular,  only when the oxygen abundance is around solar value,
there is a relatively tight relation between the 8~$\mu$m emission and
the SFR. Since the
abundance in the outer disk of M83 is about 1/5--1/10 the solar value,
we derive a relation between the 8~$\mu$m flux and the SFR using
the low oxygen abundance data points in Fig. 3 of \citet{cal07}:
\begin{equation}
SFR(M_{\bigodot}~yr^{-1})=1.6\times10^{-42}[L_{8 ~\mu m}
(erg~s^{-1})]
\end{equation}

From the 22 sources in both OuterI and OuterL that show an excess at 8~$\mu$m (and little
or no excess in the other bands) we then derive a mean SFR density for the two
regions by summing up the dust emission at 8~$\mu$m and dividing it by the total
area. By applying equation~2, we
obtain an average SFR density of
0.8$\times10^{-3}~M_{\bigodot}~yr^{-1}~kpc^{-2}$.

 The excess (over stellar photospheric) emission at 4.5~$\mu$m has been already observed
 in other galaxies, both with ISO and with \spitzer\  \citep{lu03,hel04,reg04}. This is the first time
 such excess is inferred (albeit  with large uncertainty) in the outer regions of a galaxy. The origin
 of this excess, likely due to dust emission, is still unclear.
 \citet{lu03} suggest that the excess could be due to very small grains transiently heated by single
 photons to high temperatures, $\sim$1000~K. . Our result of 14\%$\pm$6\% by flux as due to
 non--stellar emission is intermediate between the value found by \citet{reg04} for NGC7331
 (6\%) and the value found by \citet{hel04} for NGC300 (17\%) and by \citet{lu03} for a
 sample of galaxies observed with ISO.

\subsection{The Laws of Star Formation}\label{s:law}
As already documented by \citet{boi07}, the stellar and gas radial
profiles of the galaxies with extended UV emission do obey the
scaling laws of star formation \citep{ken98,ken07}. We investigate
the proportionality between the localized star formation rate
density and gas density in the outer regions of M83, by combining
the \galex\ FUV data and the 8~$\mu$m dust--only emission data with the HI
map. This analysis thus provides a verification of the {\em local} scaling laws
of star formation in the outskirts of this galaxy, in contrast with previous
results that have analyzed azimuthal averages of both the SFR and gas densities.
We will assume that the total SFR in the region is given by the sum of the SFRs
derived from the observed FUV (uncorrected for extinction) and from the 8~$\mu$m--dust
emission.  The 8~$\mu$m--emitting dust is heated by non--ionizing UV and optical photons
\citep{li02}, and can thus be adopted as a tracer of the UV photons that have been absorbed
by dust and re-emitted in the IR; therefore, the extinction--corrected UV emission will be the
sum of the observed FUV and the 8~$\mu$m--dust emission. We will use this assumption to
derive SFRs in the outskirts of M83. We note that the absence of
data on the molecular gas content is a limitation of our analysis,
which should then be interpreted with care.

At the large distances of OuterI and OuterL  from the center of the galaxy,
the response of the VLA primary beam pattern is substantially
reduced (compared to the galaxy center), the beam size is degraded, and the intrinsic $N_{HI}$
column densities are comparatively low. To control and minimize uncertainties arising
from the combination of these characteristics, we perform
photometry of regions in apertures of 20$^{\prime\prime}$ diameter, corresponding to a
physical scale of about 440~pc, slightly larger than the sizes employed so far in our analysis.
We use  newly defined regions that are measured in both the
GALEX FUV and in the dust--only 8~$\mu$m images. The larger apertures enable
us to include in each measurement multiple stellar clusters,
thus justifying the derivation of a SFR density (which would be
inappropriate for a single stellar cluster). We will also assume
that star formation has proceeded at a quasi-constant level over
the past $\approx$100~Myr, an assumption justified by the results
of the previous sections. This part of the analysis does not aim
at completeness in any sense on the selection or measurement of
 UV-- and/or 8~$\mu$m--emitting regions; it only aims at providing a range of
values for the  SFRs and HI column densities, to investigate whether star
formation at such large distances may suggest deviations from the
Schmidt--Kennicutt Law or suggest violations of the star formation
threshold \citep{mar01}. We identify a total of 54 regions
(25 in OuterI and 29 in OuterL), and obtain UV and 8~$\mu$m--dust photometry for each.

\begin{figure*}[!thb]
  \centerline{
       \epsfig{figure=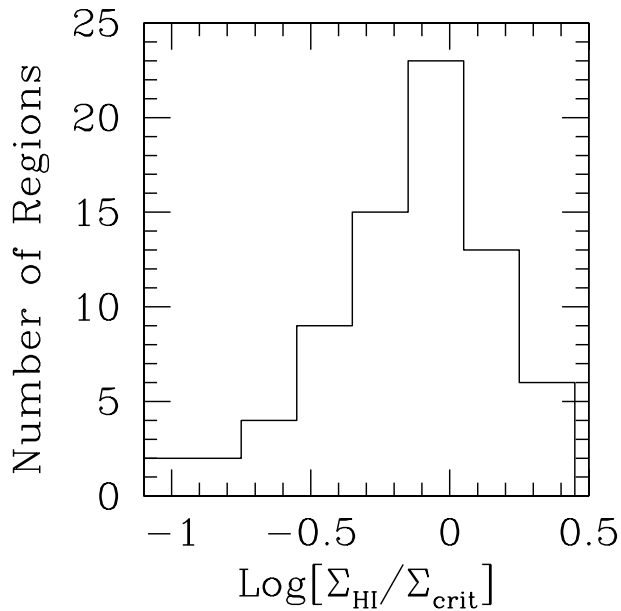,width=0.5\textwidth,angle=0}
    }
 \caption{Histogram of the ratio of the HI gas density to the critical density for
 selected UV--bright  regions in the OuterI and OuterL  fields. }
 \label{f:sigma}
 \end{figure*}

To infer the critical density for this face--on galaxy, we use the
 rotation velocity of 160~km~s$^{-1}$ measured at the distance of $\sim$20~kpc
 \citep{sof99}, assume that the velocity dispersion of the stars is negligible relative
 to the rotation velocity at this
 distance from the center, and apply equation~6 of \citet{ken89}.
 Figure~\ref{f:sigma} shows the histogram of the ratio
of the HI density to the gas critical density at the average
distance of the regions in OuterI and OuterL. As both fields are located at
$\sim$15$^{\prime}\sim$19.5~kpc distance from the center of the galaxy, we adopt a single
value of the critical density. The scatter is
large, as expected for measurements performed at low
signal-to-noise ratio, but the peak value is around
$\Sigma_{HI}$/$\Sigma_{crit}$=1.  This suggests that star formation in
these regions happens at roughly the local critical density
value. The tail towards negative values in Figure~\ref{f:sigma} may  suggest
that the HI gas is clumped over scales that are smaller than our measurement aperture
(440~pc); this is not unreasonable, since the typical HII complex has a much smaller
characteristics scale than $\sim$400~pc. However, we should stress again that the addition
of data on the molecular gas density would boost these ratios in the positive
sense.

\begin{figure*}[!thb]
  \centerline{
       \epsfig{figure=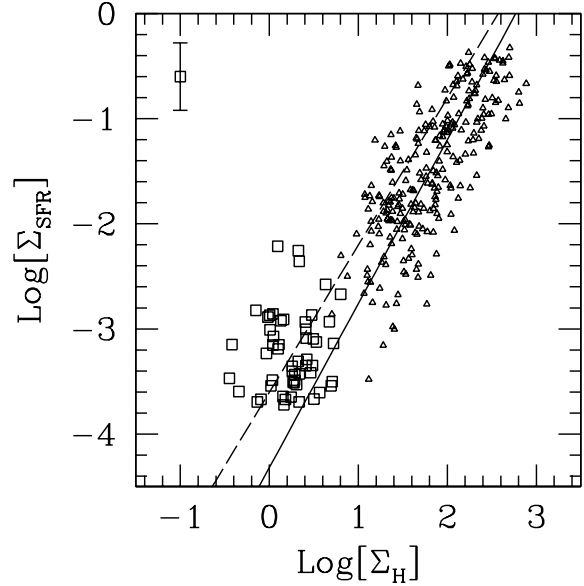,width=0.5\textwidth,angle=0}
    }
 \caption{The star formation rate density (in units of M$_{\odot}$~yr$^{-1}$~kpc$^{-2}$)
 versus gas density (in units of M$_{\odot}$~pc$^{-2}$) for  UV--bright regions (squares)
  in the OuterI and OuterL regions,
  and for regions in the disk of M51 (small triangles), as derived in \citet{ken07}. The
  typical 1~$\sigma$
  error bar is shown for the M83 data in the upper--left corner of the plot. The gas density
  for the M83 data is derived from HI alone, while it is a combination of atomic and molecular gas
  measurements for the M51 data. The continuous line is the best fit to the M51 data from \citet{ken07};
  the dashed lines is the relation from \citet{ken98} for whole galaxies.}
 \label{f:schmidt}
 \end{figure*}

The outer regions also display star formation roughly in agreement with
the Schmidt-Kennicutt Law. This is shown in
Figure~\ref{f:schmidt}, where the star formation rate density, SFRD(UV+8~$\mu$m),
versus HI gas density plot is compared
both with the mean trend observed for
$\approx$500~pc regions in M51 \citep{ken07}, and with the average trend for whole
galaxies reported in \citep{ken98}. The UV data are converted to SFRs using the formula of
\citep{ken98} modified for the Kroupa IMF. The conversion between the 8~$\mu$m--dust data and
SFR uses equation~2.  The dynamical range spanned by the M83's outskirts
data is small, and generally in the low gas density part of the
diagram, but  there is a general agreement with the trend observed for the M51 regions.
Yet  there is a significant number of M83 regions that are located above the
mean trend marked by the M51 data, implying that these data show higher SFRs for the
amount of measured gas density. We should recall that the
uncertainties in one of the two components of the SFR, i.e., the faint dust--emission--only data,
 are large (Figure~9) and the conversion from 8~$\mu$m dust
emission to SFR is highly uncertain, due to the significant dependence of the 8~$\mu$m
emission on metallicity \citep{eng05,cal07}.   In addition, the M83 data lack
measurements of molecular gas
content, which, if added, would move the data towards higher gas densities. Finally, the
observed trend is not dissimilar from what observed in low--metallicity galaxies
\citep{ken98}.

\section{Discussion and Summary}\label{s:result}
The new \spitzer\  \irac\ observations presented here, in conjunction with \galex\
data and a new HI map from THINGS,  have expanded our understanding the star
formation processes in the outskirts of M83.

The \irac\ photometry has targeted the virtually--extinction--free, mid--infrared
emission from the stars (3.6 and 4.5~$\mu$m), and the dust emission (5.8 and
8.0~$\mu$m) from stellar clusters in the extreme outer regions of M83, providing
constrains on their masses, their ages (together with the UV data), and their star
formation properties (together with the HI data).

The \spitzer\ \irac\ observations have targeted two fields, about 9.8~kpc in size at a
distance of $\sim$19.5~kpc from the center of M83. The two fields are located at about
3~times the H$\alpha$ edge in the galaxy, in correspondence of UV--emitting regions that
present a dearth of H$\alpha$ emission.  The cross-identification of UV and \irac\ sources has
yielded a final sample of 136 `bona-fide' stellar clusters (or multiple stellar clusters)
belonging to M83, after purging all potential foreground stellar and background galaxy
contaminations with a variety of methods.

Comparison between the multi--wavelength photometry and synthetic SEDs from
Starburst99 yields ages for the sources between $\sim 1$ Myr to more than 1 Gyr, which
explains the dearth of H${\alpha}$ emission in the outer regions of M83 \citep{thi05}. Our
sources show a median age around 180~Myr, which however could be a selection effect
induced by
our UV--based source selection technique. Overall, star formation has been an on--going
process in the outer regions of M83 over the past $\approx$1~Gyr. The SED fits also give
masses in the range  $10^2$ to  $10^6$ $M_{\bigodot}$ for the sources, with a median
value of $10^{4.9}~M_{\bigodot}$, comparable to the masses of globular clusters. These
results are quantitatively confirmed when the (albeit far more uncertain than our \irac\
data) 2MASS J, H, and K$_s$ data are added as constraints to the SED fitting. It should be
remarked that our sources (each encompassing a physical region of 220~pc or more in size)
could include more than one stellar cluster; this is likely true for one of the  sources in common
with the sample of HII regions of \citet{gil07}, for which those authors obtain, from optical
spectroscopy, masses and ages far lower than  what we obtain with multi--wavelength SED fitting.

The agreement between expectations and observations for the ionizing gas emission from these regions
(i.e., little emission expected overall, due to the broad range of ages shown by the stellar
clusters) appears to argue against  significant leakage of ionizing photons. This assumes
that there is little dust extinction in those regions. For significant values of the extinction, the
expected ages of the clusters would decrease, thus introducing a discrepancy between the
observed and predicted amounts of H$\alpha$ flux, and the possibility of UV photons
leakage. However, measurements of the Balmer decrement available for a few of the HII regions
\citep{gil07} support our choice of low values for the dust extinction in these regions. Presence
of ionizing photons leakage would have negligible impact on our conclusions on the star formation
law in the outer regions, because our measurements of the SFR density rely on the non--ionizing
UV stellar continuum, both as a direct measure and as an indirect measure via the 8~$\mu$m dust
emission.


Some of the sources show dust emission at 4.5~$\mu$m, 5.8~$\mu$m, and 8~$\mu$m.
We use the 8~$\mu$m dust emission to derive an average SFR density for the
two fields of 0.8 $\times10^{-3}~M_{\bigodot}~yr^{-1}~kpc^{-2}$., although this
number should be used with caution, given the many issues related to the use
of the 8~$\mu$m PAH emission for tracing star formation \citep{cal07}. The dust
emission at 4.5~$\mu$m we observe accounts for 14\%$\pm$6\% of the total flux;
this number is in between the values reported for other galaxies \citep{lu03,reg04,hel04},
although this is the first time this excess is observed in regions so far removed from the
center of a galaxy. \citet{lu03} suggest that the 4.5~$\mu$m excess could be emission from
very small dust grains  transiently heated to 1000~K. The UV photons from the our clusters
could provide the heating source for the dust, although this will need confirmation.

Star formation, thus, appears to be an unexceptional event in the outskirts of M83. From
the THINGS HI map the location of our sources is in correspondence of local HI
enhancements \citep[see also][]{thi05}, and those regions are consistent with the local
gas density to be around the critical density value, in agreement with the threshold
hypothesis of \citet{mar01}, at least when applied locally. Furthermore,  the star formation rate
densities and gas densities follows a scaling relation similar to that found for the star forming
regions in the disk of M51 \citep{ken07}. This reinforces the conclusion that the outer regions of
M83, albeit sparsely populated with stars and sites of low gas densities {\em on average},  still
form stars and stellar clusters following relations already established for the higher density
regions of disks.

\vfil \eject

\end{document}